\documentstyle{article}

\def\bigl{\left({\vrule height1.29em width0em depth1.29em}}
\def\bigr{{\vrule height1.29em width0em depth1.29em}\right)}

\def\ni{\noindent }
\def\eq #1{(\ref{#1})}          
\def\l{\left}                   
\def\r{\right}                  
\def\yx{$\{x\leftrightarrow y\}$}

\newcommand{\be}[1]{\begin{equation}\label{#1}}
\def\ee{\end{equation}}

\newcommand{\ba}[1]{\begin{array}{#1}}
\def\ea{\end{array}}

\def\fr #1#2{\frac{#1}{#2}}

\def\se #1{sec.\,\ref{#1}}

\def\y1{\mbox{$y'$}}
\def\yt{\mbox{$y''$}}

\def\fz{f_0}
\def\fo{f_1}
\def\ft{f_2}
\def\f3{f_3}

\def\fl3{\tilde{\f3}}

\def\s3t{\tilde{{s_3}}}

\def\e{{\rm e}}

\def\hyper3{{\bf hyper3}}   

\def\2F1{\mbox{$_2${F}$_1$}}
\def\1F1{\mbox{$_1${F}$_1$}}
\def\0F1{\mbox{$_0${F}$_1$}}
\def\PFQ{\mbox{$_p${F}$_q\;$}}

\begin{document}

\title{Solutions for the 
General, Confluent and Biconfluent Heun equations and their connection with Abel equations}

\author{E.S. Cheb-Terrab$^{a,b}$}

\date{}
\maketitle
\thispagestyle{empty}


\medskip
\centerline {\it $^a$CECM, Department of Mathematics}
\centerline {\it Simon Fraser University, Vancouver, British Columbia, Canada.}

\medskip
\centerline {\it $^b$Maplesoft, Waterloo Maple Inc.}


\bigskip


\maketitle

\begin{abstract}

In a recent paper, the canonical forms of a new multi-parameter class of
Abel differential equations, so-called AIR, all of whose members can be
mapped into Riccati equations, were shown to be related to the differential
equations for the hypergeometric \2F1, \1F1 and \0F1 functions. In this
paper, a connection between the AIR canonical forms and the Heun General
(GHE), Confluent (CHE) and Biconfluent (BHE) equations is presented. This
connection fixes the value of one of the Heun parameters, expresses another
one in terms of those remaining, and provides closed form solutions in terms
of \PFQ functions for the resulting GHE, CHE and BHE, respectively depending
on four, three and two irreducible parameters. This connection also turns
evident what is the relation between the Heun parameters such that the solutions
admit Liouvillian form, and suggests a mechanism for relating linear
equations with N and N-1 singularities through the canonical forms of
a non-linear equation of one order less.


\end{abstract}

\section*{Introduction}

The Heun equation \cite{ronveaux} is a second order linear equation
of the form 


\be{heun}
\yt+ \l( {\fr {\gamma}{x}}+{\fr {\delta}{x-1}}+{\fr {
\epsilon}{x-a}} \r) \y1 
+ {\fr {\alpha\, \beta\, x - q }{x \l( x-1 \r)  \l( x-a \r) }}\, y = 0,
\end{equation}

\ni where $\{\alpha, \beta, \gamma, \delta, \epsilon, a, q\}$ are constant
with respect to $x$, are related by
$\gamma+\delta+\epsilon=\alpha+\beta+1$, and $a \neq 0, a\neq 1$. This
equation has four regular singular points, at $\{0, 1, a, \infty\}$. Through
confluence processes, equation \eq{heun}, herein called the General Heun
Equation (GHE), transforms into four other multi-parameter equations
\cite{decarreau}, so-called Confluent (CHE), Biconfluent (BHE),
Doubleconfluent (DHE) and Triconfluent (THE). Through transformations of the
form $ y \rightarrow P(x)\,y$, these five equations can be written in normal
form\footnote{A second order linear ODE is in {\em normal form} when the
coefficient of $\y1$ is equal to zero - see the Appendix.}, using the notation of \cite{decarreau}, in terms of arbitrary
constants $\{a,\, A,B,C,D,E,F\}$; for the
6-parameter GHE \eq{heun} we have



\be{GHE_NF}
\yt+ \l( {\fr {A}{x}} + {\fr {B}{x-1}} -{\fr {A+B}{x-a}} 
+ {\fr {D}{{x}^{2}}} + {\fr {E}{ \l( x-1 \r) ^{2}}} + {\fr {F}{
\l( x-a \r) ^{2}}} \r) y = 0
\ee

\ni The 5-parameter CHE,


\be{CHE_NF}
\yt+ \l( A+{\fr {B}{x}}+{\fr {C}{x-1}}+{\fr {D}{{x}^{2}}}+{
\fr {E}{ \l( x-1 \r) ^{2}}} \r) y  =  0,
\ee

\ni has two regular singularities at $\{0,1\}$ and one irregular singularity
at $\infty$. The 4-parameter BHE,


\be{BHE_NF}
\yt+ \l( -{x}^{2}+B\,x+C+{\fr {D}{x}}+{\fr {E}{{x}^{2}}} \r) y  =  0,
\ee

\ni has one regular singularity at $0$ and one irregular singularity at
$\infty$. The 4-parameter DHE,


\be{DHE_NF}
\yt+ \l( A+{\fr {B}{x}}+{\fr {C}{{x}^{2}}}+{\fr {D}{{x}^{3}}}
+{\fr {A}{{x}^{4}}} \r) y  =  0,
\ee

\ni has two irregular singularities at $\{0,\infty\}$. The 3-parameter THE,
with one irregular singularity at $\infty$, is


\be{THE_NF}
\yt+ \l( -\fr{9}{4}\,{x}^{4} + C\,{x}^{2} + D\,x + E \r) y  =  0
\ee


Eq.\eq{heun}, originally studied by Heun as a generalization of Gauss'
hypergeometric ($_p{\rm F}_q$) equation, as well as these related confluent families
represented by (\ref{CHE_NF}--\ref{THE_NF}), appear in applications
in varied areas\footnote{For a list of applications of Heun's equations
compiled in 1995 see p.340 of \cite{ronveaux}.}. As a sample of recent
related works, in \cite{starinets} quase-normal modes of near extremal black
branes are found solving a singular boundary value problem for \eq{heun}; in
\cite{tolstikhin}, hyper-spherical harmonics, with applications in 
three-body systems, are developed in connection with the solutions of
\eq{heun}; in \cite{anguelova}, a method of calculation of propagators for
the case of a massive spin 3/2 field, for arbitrary space-time dimensions
and mass, is developed in terms of the solutions of \eq{heun}; in
\cite{finkel}, parametric resonance after inflation is discussed in
connection with the solutions of a particular form of \eq{heun}. The
separation of variables for the Schr\"odinger equation in a large
number of problems results in Heun type equations too, typically
for the radial coordinate, and also non-linear formulations involving
Painlev\'e type equations \cite{slavyanov} can be derived from Heun
equations regarded as quantum Hamiltonians. A number of traditional
equations of mathematical physics, as for instance the Lam\'e, spheroidal
wave, and Mathieu equations, are also particular cases of Heun equations.

The solutions for these five Heun equations are the subject of current study
\cite{chazy}--\cite{craster2002}. In this paper, a hitherto unknown connection
between Heun equations and a single multiparameter Abel equation \cite{abel2}, known to
have canonical forms solvable in terms of \PFQ functions \cite{abel3}, is shown. This
connection fixes one of the parameters and expresses another one in terms of those
remaining in the Heun equations \eq{GHE_NF},
\eq{CHE_NF} and \eq{BHE_NF}, and provides exact closed form solutions for
the resulting non-trivial 4-parameter GHE, 3-parameter CHE, and 2-parameter BHE families. The
solutions are linear combinations involving \2F1 or \1F1 functions, and this
connection with Abel equations also turns evident what is the relation between the
Heun parameters such that the solutions of these three families admit
Liouvillian form. From these results, an alternative approach to finding the same
solutions, by exploring non-local transformations, is derived, and some of
these solutions are shown to match those derived in \cite{craster2002} using
an essentially different approach. 

The multiparameter BHE, CHE and GHE equations solved in this paper are non-trivial 
in that they are irreducible, not degenerate, cases:
the number of their singularities cannot be reduced and 
the equations cannot be mapped into \PFQ equations through extended
transformations of the form

\be{tr_hyper3}
x \rightarrow \fr{\alpha x^k+\beta}{\gamma x^k+\delta}, 
\ \ \ y \rightarrow P(x)\, y
\ee

\ni where $\{\alpha, \beta, \gamma, \delta, k\}$ are constants and
$P(x)$ is an arbitrary Liouvillian function. Hence, the solutions being
presented cannot be reduced to \PFQ solutions of the form

\be{hyper3_sol}
y = P(x)\, \PFQ\l(..., ...,  \fr{\alpha x^k+\beta}{\gamma x^k+\delta}\r).
\ee

\ni and it is in this extended sense that, herein, we say the equations
being solved are not of \PFQ type\footnote{There exist symbolic computation
libraries that systematically resolve the equivalence of linear ODEs under
\eq{tr_hyper3} - see \cite{hyper3}.}.

Apart from being a way to relate $\mbox{Heun} \leftrightarrow
\mbox{\PFQ}$equations, leading to solutions to the former ones, this
connection $\mbox{Heun} \leftrightarrow \mbox{Abel}$ is important in itself:
Abel equations also appear frequently in applications
\cite{winternitz}--\cite{sachdev} and through this connection it is possible to study their
properties by studying those of the related linear equations. It is
implicit in the existence of this $\mbox{Heun} \leftrightarrow \mbox{Abel}$ link that there exists an
equivalent link, between linear equations of order $n$ with $N$ and $N-1$
singularities and related confluent cases, through single non-linear ``Abel-like" equations of
order $n-1$.

The paper is organized as follows. In \se{Theory}, some results of \cite{abel2} and \cite{abel3} are reviewed
and a connection between Heun and Abel equations is made
explicit. In \se{BHE}, \ref{CHE} and \ref{GHE}, the restrictions that this connection implies
on the parameters entering the BHE, CHE and GHE equations are derived, and
it is shown how a sequence mapping $\mbox{Heun} \rightarrow \mbox{Abel}
\rightarrow \PFQ$ equations can be composed to obtain transformations
mapping $\mbox{Heun} \rightarrow \PFQ$ equations. When the aforementioned
restriction on the Heun parameters holds, these transformations lead to
closed form solutions for the GHE \eq{GHE_NF}, CHE
\eq{CHE_NF} and BHE \eq{BHE_NF}, expressed in terms of exponentials of
integrals of \2F1 or \1F1 functions. In \se{alternative}, taking advantage of
the results of the previous sections, an alternative derivation is developed
leading to solutions free of integrals. A discussion around these results is
found in \se{comparison} and \ref{discussion}. In an Appendix, the formulas relating the normal
and canonical forms for the Heun equations are included for completeness, as
well as symbolic computation input permitting the verification of the
solutions presented.


\section{A connection between Heun, Abel and \PFQ differential equations}
\label{Theory}

The transformations being presented, relating Heun and \PFQ hypergeometric
equations, were obtained by composing transformations which
map Heun, Riccati, Abel and \PFQ equations according to the
sequence

$$
Heun \rightarrow Riccati \rightarrow Abel \rightarrow Abel_{canonical} \rightarrow Riccati \rightarrow \PFQ
$$

\ni As we shall see in \se{alternative}, knowing the form of these
transformations, one can re-derive them in an alternative way, shortcutting
the step which goes through Abel equations. In this section, however, the $Heun
\leftrightarrow Abel$ connection is kept visible: Abel equations are
relevant by themselves and it was through this connection that
the $Heun \leftrightarrow \PFQ$ relation being presented became evident. 

Abel equations of the second kind \cite{kamke}  are equations of the form 

\be{abel_2k}
\y1 = \fr{\f3(x)\, y^3 + \ft(x)\, y^2 + \fo(x)\, y + \fz(x)}{g_1(x)\,y + g_0(x)},
\end{equation}

\ni where the $\{f_i,\ g_i\}$ are arbitrary functions and either $f_3(x)
\neq 0$ or $g_1(x) \neq 0$. In \cite{abel2} it is shown that, departing from
a Riccati type equation,

\be{riccati}
\y1= h_2(x)\, y^2 + h_1(x)\, y + h_0(x),
\end{equation}

\ni by suitably restricting the form of the mappings $h_i$ and making use of
the {\em inverse} transformation \yx\footnote{By \yx\ we mean changing
variables $\{x = u(t),\, y(x) = t\}$ followed by
renaming $\{u = y,\, t = x\}$.}, one can construct an Abel equation,

\be{AIR_rho}
\y1={\frac{(y-\rho_1) (y-\rho_2) (y-\rho_3)}
{ \l( s_{2}\,x^2 + s_{1}\,x + s_{0} \r) y + r_{2}\,x^2 + r_{1}\,x + r_{0}}}
\ee

\ni \ni where the $\{s_i, r_i, \rho_i\}$ are constants, and $s_2 \neq 0$ or $r_2
\neq 0$. This Abel equation is representative of a multi-parameter class
all of whose members can be transformed into Riccati
equations \eq{riccati} using \yx, and from there into second order linear equations using the
$Riccati \rightarrow linear$ mapping \cite{kamke}

\be{riccati_to_linear}
y \rightarrow -{\fr {\y1}{h_{{2}} ( x )\, y}}
\ee

\ni The equations of this Abel class, named ``Abel
Inverse Riccati" (AIR) in \cite{abel2}, are then generated from \eq{AIR_rho}

\ni by applying to it class transformations of the form

\be{tr_2k}
\{x \rightarrow F(x),\ \ y \rightarrow \fr{P_1(x)\,  y + Q_1(x)}{P_2(x)\, y + Q_2(x)}\},
\ee


\ni where $\{F,\,P_1,\,P_2,\,Q_1,\,Q_2\}$ are arbitrary mappings with
$F' \neq 0,\ P_1\,Q_2 - P_2\,Q_1 \neq 0$. The relevance of the AIR class can
be inferred from the fact that most of the Abel solvable equations found in
the literature\footnote{For a collection of these see \cite{abel1}.} are
shown in \cite{abel2} to be particular members of AIR. 

An important property of \eq{AIR_rho} is that its connection with second
order linear equations, that is, its ``Inverse Riccati" character, is
invariant under M\"obius (linear fractional) changes of $x$ and $y$. This
property is used in \cite{abel3} to accomplish a full classification of
\eq{AIR_rho} in terms of six canonical forms. For that purpose, through
M\"obius changes of $y$, \eq{AIR_rho} is first transformed into

\be{can_forms}
\y1={\fr {P \l( y \r) }{ \l( s_{2}\,x^2+s_{1}\,x+s_{0} \r) y + r_{2}\,x^2 + r_{1}\,x+r_{0}}}
\ee

\ni for some new constants $\{s_i\,r_i\}$, with $P(y)$ equal to $y\,(y-1),\,
y$ or $1$, respectively according to whether in \eq{AIR_rho} there are
three, two or only one distinct roots $\rho_i$. As shown in \cite{abel3},
each of these three cases splits further into two subcases, and the six resulting canonical forms
are solvable in terms of \2F1, \1F1 and \0F1 functions; in this way, closed form
\PFQ solutions can be constructed for the whole AIR class.

The key observation now is that the AIR equation \eq{can_forms} is also
connected in a surprisingly simple manner to the Heun family of equations.
As we shall see, by applying to \eq{can_forms} the transformation \yx, one
obtains a Riccati equation, and by transforming it further into a second
order linear equation using \eq{riccati_to_linear}, one directly obtains the
GHE, CHE or BHE Heun equations (with some restrictions on the parameters),
respectively according to the three possible values of $P(y)$. Since the AIR
\eq{can_forms} admits solutions expressible using \PFQ functions for the
three possible values of $P(y)$, the GHE, CHE and BHE Heun families which
can respectively be derived from \eq{can_forms} also admit closed form solutions
expressible in terms of these \PFQ functions.

\section{Closed form solutions for a subfamily of the BHE}
\label{BHE}

Considering first the simplest case, where the three roots $\rho_i$ in
\eq{AIR_rho} are equal, in \eq{can_forms} we have $P(y) = 1$
and so the AIR equation becomes

\be{AIR1}
\y1={\fr {1}{ \l( s_{2}\,x^2+s_{1}\,x+s_{0} \r) y + r_{2}\,x^2 + r_{1}\,x+r_{0}}}
\ee

\ni Recalling that either $s_2 \neq 0$ or $r_2 \neq 0$, changing variables using \yx\ we obtain the Riccati form

\begin{equation}
\y1={{ \l( s_{2}\,x + r_{2} \r)\,{y}^{2} + { \l( s_{1}\,x+r_{1} \r)\, y } + {s_{0}\,x+r_{0}} }}
\ee

\ni Using \eq{riccati_to_linear}, this equation is transformed into
the second order linear equation

\be{AL_10F1}
\yt={\fr 
    { \l( s_{2}\,s_{1}{x}^{2}+ \l( s_{2}r_{1}+ s_{1}r_{2} \r) x + s_{2} + r_{2}r_{1} \r)}
    {s_{2}\,x + r_{2}}}\,\y1
- \l( s_{2}s_{0}{x}^{2} + \l( s_{2}r_{0} + s_{0}r_{2} \r) x + r_{2}r_{0} \r) y
\ee

\ni This equation has one regular singularity at $-r_2/s_2$, one irregular
singularity at $\infty$, and, by rewriting it in normal form, it is straightforward to
verify that it is the BHE equation \eq{BHE_NF} with one of its four
parameters fixed and two other ones interrelated. For that purpose, we note first that the case $s_2 = 0$ presents no
interest since it directly simplifies \eq{AL_10F1} to a \PFQ equation.
Assuming $s_2 \neq 0$ in \eq{can_forms}, we take $s_2 = 1$
without loss of generality. To have the regular singularity of \eq{AL_10F1}
located at $0$, it suffices to take $r_2 = 0$, and, taking $s_0 = s_1^2/4\, -
1$, the coefficient of $x^{2}$ in the normal form of \eq{AL_10F1} will be as in
\eq{BHE_NF}. In summary, using $y \rightarrow 
\sqrt {x}\,{\exp(({x \l( s_{{1}}\,x + 2\,r_{{1}} \r)/4\, })}\, y$ 
to rewrite \eq{AL_10F1} in normal form
at $\{s_2=1, r_2=0, s_0 = s_1^2/4\, - 1\}$, the equation becomes

\be{AL_10F1_NF_0r2}
\yt= \l( {x}^{2}+ \l( \fr{s_{{1}}r_{{1}}}{2} -r_{{0}} \r) x
+ \fr{{r_{{1}}}^{2}}{4} + {\fr {r_{{1}}}{2\,x}}+ \fr{3}{4\,{x}^{2}} \r) y
\ee

\ni which is the BHE \eq{BHE_NF} at
$\{B=r_{{0}}-s_{{1}}r_{{1}}/2,\, D=-r_{{1}}/2,\, C = -D^2,\, E=-3/4\}$. 

The relevance of this result is in that, on the one hand,
\eq{AL_10F1_NF_0r2} is a non-trivial 2-parameter form of the BHE for which solutions are
not known in general; on the other hand, as shown in \cite{abel3}, the Abel
equation \eq{AIR1}, from which \eq{AL_10F1_NF_0r2} is derived, is solvable in terms of \1F1 and \0F1 (Kummer and Bessel) hypergeometric
functions. Therefore, a closed form solution for the BHE \eq{AL_10F1_NF_0r2}
can also be expressed using these \PFQ functions. 

It is interesting, when possible, to separate the Liouvillian from the
Non-Liouvillian solutions of \eq{AL_10F1_NF_0r2}, since the former ones represent the ``special 
cases", where the solution is representable in terms of known functions. 
For that purpose, the parameters $\{r_0,r_1\}$ are redefined in terms of new
parameters $\{\sigma,\,\tau\}$ according to\footnote{The motivation for this
particular choice of $\{\sigma,\tau \}$ becomes clear below, in connection
with the form of equation \eq{AIR_1_canonical}.} 

\be{sigma_tau}
r_{{0}}=-2\,\sigma+s_{{1}}\tau,\ \ r_{{1}}=2\,\tau
\ee

\ni With this choice, the BHE \eq{AL_10F1_NF_0r2} also becomes an equation
explicitly depending on only two parameters $\{\sigma,\tau\}$,


\be{BHE_1}
\yt= \l( {x}^{2}+2\,\sigma\,x+{\tau}^{2} + {\fr {\tau}{x}} + \fr{3}{4\,x^{2}} \r) y,
\ee

\ni and \eq{AIR1} at $\{s_2=1, r_2=0, s_0 = s_1^2/4\, - 1\}$ becomes

\be{AH_1}
\y1= \fr{1} {\l( {x}^{2}+s_{{1}}x + \l( s_{{1}}+2 \r) \l( s_{{1}}-2 \r)/4 \r)\, y
+ 2\,\tau\,x - 2\,\sigma+s_{{1}}\tau}
\ee

\subsection{Liouvillian solutions for the BHE \eq{BHE_1} when $\sigma = \pm \tau$}
\label{Liouvillian_BHE}

\ni As explained in \se{solution_to_BHE}, when $\sigma = \pm \tau$, the BHE
\eq{BHE_1} can be obtained from an equation of the form $\yt + J(x)\, \y1 =
0$, ``missing $y$", through a Liouvillian transformation, and so it admits Liouvillian
solutions, computable using the relatively new Kovacic algorithm \cite{kovacic}. Concretely, for
$\sigma = \tau$, \eq{BHE_1} becomes

\be{BHE_sigma_equal_tau}
\yt= \l( {x}^{2}  + {\tau}^{2} + \tau\l(2\,x + {\fr {1}{x}}\r) + \fr{3}{4\,x^{2}} \r) y,
\ee

\ni the aforementioned equation ``missing $y$" is

\be{BHE_missing_y}
\yt={\fr { 2\,x\,({x}+\tau)+1 }{x}}\,\y1,
\ee

\ni and the transformation mapping this equation into \eq{BHE_sigma_equal_tau} is 
\be{tr_to_missing_y}
y \rightarrow \sqrt {x}\,{\e^{(x(x + 2\,\tau)/2)}}\,y
\ee

\ni Hence a general solution for \eq{BHE_sigma_equal_tau} is\footnote{In
\eq{BHE_sol_sigma_equal_tau} and \eq{BHE_sol_sigma_equal_minus_tau}, {erf} and
{erfi} are respectively the error and imaginary error functions - see
\cite{abramowitz}.}

\be{BHE_sol_sigma_equal_tau}
y={\fr {{\e^{-x \l( x+2\,\tau \r)/2 }}}{\sqrt {x}}}\,C_1
+ {\fr {\sqrt {\pi }\,{\e^{x \l( x+2\,\tau \r)/2 }}
        -\pi \,\tau\,{\rm erfi} ( x+\tau)\,
        {\e^{-x \l( x+2\,\tau \r)/2 -{\tau}^{2}}}}{\sqrt{x}}}\,C_2
\ee

It is important to note that computing the transformation \eq{tr_to_missing_y}, which maps an equation
in normal form, like \eq{BHE_sigma_equal_tau}, into one that is ``missing y", like
\eq{BHE_missing_y}, is not a trivial operation, and is entirely equivalent
to computing the solution for \eq{BHE_sigma_equal_tau}. These mappings are formally performed with 
the aid
of infinitesimal symmetry generators, and in the case of linear ODEs, the
computation of these infinitesimals
indeed requires solving the ODE itself \cite{stephani}.
Besides the power of Kovacic's
algorithm, which computes these solutions systematically just by assuming the solution 
field (Liouvillian), it is also remarkable that the condition $\sigma = \pm \tau$ for
the existence of these Liouvillian solutions of the BHE \eq{BHE_1} is directly
evident in the canonical form of the corresponding Abel equation
\eq{AIR_1_canonical} below.

Finally, when in \eq{BHE_1}, $\sigma = -\tau$, the same treatment with Kovacic's
algorithm results in the general solution

\be{BHE_sol_sigma_equal_minus_tau}
y = {\fr {{\e^{x \l( x-2\,\tau \r)/2 }}}{\sqrt {x}}}\,C_1
+ {\fr {\sqrt {\pi }\,{\e^{-x \l( x-2\,\tau \r)/2 }}
    -\pi\, \tau\,{{\rm erf}(x-\tau)}\,
    {\e^{x \l( x-2\,\tau \r)/2 +{\tau}^{2}}}}{\sqrt {x}}}\,C_2
\ee

\subsection{A solution in terms of \1F1 functions for the BHE \eq{BHE_1} when $\sigma^2 \neq \tau^2$}
\label{solution_to_BHE}

A transformation relating \eq{BHE_1} to a \1F1 differential equation, 
providing a solution for \eq{BHE_1} when $\sigma^2 \neq \tau^2$, is
constructed by composing three transformations: one which maps the BHE
\eq{BHE_1} into the AIR equation \eq{AH_1}; one which maps \eq{AH_1} into an AIR
equation admitting \1F1 solutions; and finally, one which maps that AIR equation
into a \1F1 equation.

Reversing the transformations used to derive \eq{AL_10F1_NF_0r2} from
\eq{AIR1}, the transformation mapping the Heun equation \eq{BHE_1} into the
Abel AIR \eq{AH_1} is

\be{TR1_1}
\l\{
x \rightarrow y,
\ \ 
y \rightarrow {\fr {{\e^{-\l( \int \!x\,y\,y'\,{dx} + \fr{s_{{1}}}{4}{y}^{2} + \tau\,y \r) }}}{\sqrt {y}}}
\r\}
\ee

\ni According to \cite{abel3}, the transformation mapping \eq{AH_1} into a
canonical form of AIR admitting a \PFQ solution is

\be{TR2_1_cumbersome}
\l\{
x \rightarrow  \l( {\fr {\sqrt {2}\,x}{2\,(\tau + \sigma)}}+\fr{1}{2}\r)^{-1} - \fr{s_{{1}}}{2}-1,
\ \ 
y \rightarrow \fr{\sqrt {2}\, y}{2} - \sigma
\r\}
\ee

\ni The resulting AIR canonical form is

\be{AIR_1_canonical}
\y1= \fr{1}
{x\,y + {x}^{2}+ \l( {\sigma}^{2}-{\tau}^{2} \r)/2}
\ee

\ni This form turns evident the motivation for introducing $\{\sigma,\tau\}$
according to \eq{sigma_tau}. For $\sigma = \pm \tau$, the independent term in
the denominator of \eq{AIR_1_canonical} cancels, and hence, when
transforming this equation into a second order
linear equation, we will obtain one of the form $\yt + J(x)\,\y1 = 0$ with rational $J(x)$,
admitting a constant for solution. Since the transformation of such an
equation into a normal form like \eq{BHE_1} is Liouvillian, the normal form
of the equation will admit Liouvillian solutions.

After having determined a condition for the existence
of Liouvillian solutions, for the purpose of relating the Heun
equation \eq{BHE_1} to a \PFQ equation, a simpler derivation is possible if
instead of using \eq{TR2_1_cumbersome} we use

\be{TR2_1}
x \rightarrow \fr{1}{x}- \fr{s_{{1}}}{2}-1,
\ee

\ni which does not lead to the canonical form \eq{AIR_1_canonical}, but
still leads to an AIR equation admitting \PFQ solutions

\be{AIR_BHE_alternative}
\y1= \fr{1}{( 2\,x -1)\, y + 2\,( \sigma\ + \tau )\,x^2 - 2\,\tau\,x }
\ee

\ni Following \cite{abel3}, this equation is transformed into a Riccati equation, then
into a linear second order one using

\be{TR3_1}
\l\{
x \rightarrow y,
\ \ 
y \rightarrow -{\fr {\y1}{ 2 \l( \tau + \sigma \r) y}}
\r\}
\ee

\ni leading to


\be{1F1_BHE}
\yt= 2 \l( x - \tau \r) \y1 + 2 \l( \tau+\sigma \r) x\,y
\ee

\ni Finally, using $\{ x \rightarrow \l( x+\sigma \r) ^{2},\ y
\rightarrow {{( \l( x+\sigma \r) {\e^{-x \l( \tau+\sigma \r)
}})^{-1}}\,y}\}$, equation \eq{1F1_BHE} is obtained from the confluent
\1F1 hypergeometric equation

\be{1F1_mu_nu}
x\,\yt + (\nu - x)\, \y1 - \mu\, y = 0
\ee

\ni at $\{\mu = (2 + {\tau}^{2} - {\sigma}^{2})/4,\ \nu=3/2\}$, from where
the solution to \eq{1F1_BHE}, in terms of the Kummer M and U
functions\footnote{An equivalent form of this solution in terms of \1F1 functions is
$$
y={\e^{-x \l( \tau+\sigma \r) }} 
\l( 
{\1F1\l(({\tau}^{2}-{\sigma}^{2})/4 ;\,1/2;\, \l( x+\sigma \r) ^{2}\r)}\,C_1 
+ \l( x+\sigma \r){\1F1\l(1/2 + ({\tau}^{2}-{\sigma}^{2})/4;\,3/2;\, \l( x+\sigma \r) ^{2}\r)} \,C_2 
\r) 
$$.
}  \cite{abramowitz}, is

\be{ans_PFQ_1}
y = {\e^{-x \l( \tau+\sigma \r) }}\l(
{{\rm M}\l(\fr{{\tau}^{2}-{\sigma}^{2}}{4},\fr{1}{2}, \l( x+\sigma \r)^{2}\r)}\,C_1 
+ {{\rm U}\l(\fr{{\tau}^{2}-{\sigma}^{2}}{4},\fr{1}{2}, \l( x+\sigma \r)^{2}\r)}\,C_2 \r)
\ee

Summarizing, we depart from the Heun Biconfluent equation \eq{BHE_1} and
arrive at the \1F1 equation \eq{1F1_BHE} with solution \eq{ans_PFQ_1} through a process of the
form

$$
Heun \rightarrow Abel \rightarrow Abel_{_1F_1 solvable} \rightarrow \1F1
$$

\ni The three transformations used, \eq{TR1_1}, \eq{TR2_1} and \eq{TR3_1},
can be combined into a single transformation, mapping the BHE
\eq{BHE_1} into the \1F1 \eq{1F1_BHE} in one step:

\be{tr_tot}
y \rightarrow
\fr{1}{\sqrt {x}}\ {\exp\l({\fr {{x}^{2}}{2} - \tau x + 2 \l( \tau+\sigma \r) \int \!{\fr {x\,y}{y'}}\,{dx}}\r)}
\ee

\ni Therefore, a closed form solution for the BHE \eq{BHE_1} when $\sigma^2
\neq \tau^2$ is given by this transformation \eq{tr_tot}, where in the
``right-hand-side" the value of $y$ is given by \eq{ans_PFQ_1}. By exploring
some properties of linear differential equations discussed in \se{alternative}, it is
possible to express the solution \eq{tr_tot} as a linear combination of
\1F1 functions with non-constant coefficients and entirely free of integrals
- see \eq{tr_tot_1_free_of_integrals}.

An independent verification of \eq{tr_tot}, or its form
free of integrals \eq{tr_tot_1_free_of_integrals}, as well as of the
Liouvillian solutions \eq{BHE_sol_sigma_equal_tau} and
\eq{BHE_sol_sigma_equal_minus_tau}, was performed in the Maple symbolic
computation system - see the Appendix.


\section{Closed form solutions for a subfamily of the CHE}
\label{CHE}

A confluent family of Heun equations can be derived from \eq{AIR_rho} when,
among the three roots $\rho_i$, only two are different. Hence, in
\eq{can_forms} we have $P(y) = y$ and the starting AIR equation is

\be{AIR2_draft}
\y1={\fr {y}{ \l( s_{2}\,x^2+s_{1}\,x+s_{0} \r) y + r_{2}\,x^2 + r_{1}\,x+r_{0}}}
\ee

\ni As in the previous section, changing variables using \yx to obtain a
Riccati type equation, then using \eq{riccati_to_linear}, we obtain the linear equation

\be{AL_1F1_draft}
\yt=
{\fr 
    {\l( s_{{2}}s_{{1}}{x}^{2} + \l( s_{{1}}r_{{2}} + s_{{2}}r_{{1}}\r) x - r_{{2}} + r_{{2}}r_{{1}} \r)}
        {x \l( s_{{2}}x + r_{{2 }} \r) }}\, \y1
- {\frac
        { \l( s_{{2}}s_{{0}}{x}^{2}+ \l( s_{{2}}r_{{0}} + s_{{0}}r_{{2}} \r) x + r_{{2}}r_{{0}} \r)}
        {{x}^{2}}
    } \,y
\ee

\ni This equation has two regular singularities at $\{0,-r_2/s_2\}$ and one
irregular singularity at $\infty$, and by rewriting it in normal form, its confluent
Heun type becomes evident. As shown below, the implicit restrictions in \eq{AL_1F1_draft} as compared to the most
general case \eq{CHE_NF} consist of having one of the five parameters, $E$,
fixed and another one, $B$, being a function of the remaining three.

To derive the relation between the parameters of the CHE equations
\eq{CHE_NF} and \eq{AL_1F1_draft}, and then a solution for \eq{AL_1F1_draft}
in the non trivial cases, we start by noting that, when $s_2 = 0$,
\eq{AL_1F1_draft} simplifies to a \PFQ equation; the interesting case is
$s_2 \neq 0$, which is equivalent to taking $s_2 = 1$ in \eq{AIR2_draft}.
The regular singularities of \eq{AL_1F1_draft} are fixed at $\{0,1\}$ by
taking $r_2 = -1$, and the term independent of $x$ in the normal form of
\eq{AL_1F1_draft} is fixed to be $A$, as in \eq{CHE_NF}, by taking $s_0 =
s_1^2/4 + A$. In summary, rewriting \eq{AL_10F1} in normal form, using $y
\rightarrow {x}^{(r_{{1}}-1)/2}\,\sqrt{x-1}\,{\exp({s_{{1}}\,x/2})} \,y$, at
$\{s_2=1, r_2=-1, s_0 = s_1^2/4 + A\}$, the equation becomes


\be{CHE_NF_PFQ}
\yt = \l( 
    - A 
    + {\fr 
        {{s_{{1}}}^{2}/2+ s_{{1}}( r_{{1}}-1) - r_{{1}} - 2\,r_{{0}} + 2A + 1}
        {2\,x}}
    + {\fr {s_{{1}} + r_{{1}} - 1}{2\,(x-1)}}
    + {\fr 
        {{r_{{1}}}^{2}+4\,r_{{0}}-1}
        {4\,{x}^{2}}}
    + \fr{3}{4\, \l( x-1 \r)^{2}}
    \r) y
\ee

\ni which is the CHE \eq{CHE_NF} at

\begin{equation}
B=-A-D-{C}^{2},\ \ C= \fr{1-s_{{1}}-
r_{{1}}}{2},\ \ D= \fr{1-{r_{{1}}}^{2}}{4} -r_{{0}},\ \ E=-\fr{3}{4}
\ee

As shown in \cite{abel3}, the Abel AIR equation \eq{AIR2_draft} can always
be solved in terms of \1F1 hypergeometric functions, from where a closed
form solution expressed using \1F1 can also be constructed for the CHE
\eq{CHE_NF_PFQ}. In order to separate the Liouvillian special cases of the solutions of
\eq{CHE_NF_PFQ} from the generally non-Liouvillian ones, the parameters $\{r_0,r_1\}$ are redefined in terms of new
parameters $\{\sigma,\tau\}$ according to $ \{r_{{0}}= ( 1-2\sigma)\,{p}^{2} -
{s_{{1}}}^{2}/4 + s_{{1}}\tau p,\,r_{{1}}=2\,\tau p-s_{{1}} \}$. This
redefinition is derived as in the previous section, from the canonical form
of the AIR \eq{AH_2} (see comments after \eq{AIR_1_canonical} and also \cite{abel2}).
Introducing also $A = -\lambda^2$ (see \cite{ronveaux} p.94), \eq{CHE_NF_PFQ}
becomes


\be{CHE_2}
\yt = 
\l( 
    {\lambda}^{2} 
    + {\fr {2\,( \sigma - 1) {\lambda}^{2} - \tau\,\lambda + 1/2}{x}}
    + {\fr {\tau \lambda - 1/2}{x-1}}
    + {\fr {( {\tau}^{2} -2\,\sigma + 1 )\,{\lambda}^{2} - 1/4}{{x}^{2}}}
    + \fr{3}{4\, \l( x-1 \r)^{2}}
    \r) y
\ee

\ni that is, a 3-parameter equation, and the AIR equation \eq{AIR2_draft} becomes


\be{AH_2}
\y1 = 
    {\fr {y}
    { \l( {x}^{2} + s_{{1}}x - {\lambda}^{2} + {s_{{1}}}^{2}/4 \r) y
    - {x}^{2}
    + ( 2\,\tau \lambda - s_{{1}})\, x 
    + ( 1 -2\,\sigma )\, {\lambda}^{2}
    + s_{{1}}\tau \lambda
    - {s_{{1}}}^{2}/4}
    }
\ee

\subsection{Liouvillian solutions for the CHE \eq{CHE_2} when $\sigma = \pm \tau$}
\label{Liouvillian_CHE}

\ni As in the previous section, we know, by construction, that at $\sigma =
\pm \tau$, the CHE \eq{CHE_2} admits Liouvillian solutions. Using Kovacic's
algorithm, for $\sigma = \tau$, 


\begin{equation}
y={\fr 
    {{x}^{\l( 1-\tau \r) \lambda + 1/2}\, {\e^{-\lambda\,x}}}
    {\sqrt {x-1}}}
    \l( C_1 +  
    \l( \Gamma( 2 \l( \tau-1\r)\lambda + 1,\, -2\,\lambda\,\, x) + 2\,\lambda\, \Gamma( 2 \l( \tau-1 \r)\lambda ,\,-2\,\lambda\,x)  \r) C_2 
    \r)    
\ee

\ni where $\Gamma$ (of two arguments - see (6.5.3) in \cite{abramowitz}) is the incomplete gamma function.
For $\sigma = -\tau$, 


\begin{equation}
y={\fr 
    {{x}^{-\l(1+\tau \r)\lambda+1/2 }\, {\e^{\lambda\,x}}}
    {\sqrt {x-1}}}
    \l(
        C_1
        +  \l( 2\,\lambda\,\Gamma\l( 2\l( \tau+1 \r) \lambda,\, 2\,\lambda\,x \r) 
            -\Gamma\l( 2\l( \tau+1 \r)\lambda+1,\, 2\,\lambda\,x \r)  \r) C_2 \r)    
\ee

\subsection{A solution in terms of \1F1 functions for the CHE \eq{CHE_2} when $\sigma^2 \neq \tau^2$}

As in the previous section, a solution to \eq{CHE_2} when $\sigma^2 \neq
\tau^2$ is constructed by composing three transformations: the one which maps the CHE
\eq{CHE_2} into the AIR \eq{AH_2}; one which maps \eq{AH_2} into an AIR
equation admitting \1F1 solutions; finally, one which maps that AIR equation
into a \1F1 equation.

Reversing the transformations used to derive \eq{CHE_NF_PFQ} from
\eq{AIR2_draft}, the transformation mapping the Heun equation
\eq{CHE_NF_PFQ} into the Abel AIR \eq{AH_2} is


\be{TR1_2}
\l\{
x \rightarrow y,
\ \ 
y \rightarrow {\exp\l({-\int \!{\fr {
    \l(
        \l( x + s_{{1}}/2 \r) {y}^{2}
        - \l( 2\,x - \tau\,\lambda + s_{{1}} \r) y
        + x - \tau\,\lambda + (s_{{1}} + 1)/2
    \r) y'}
    { y \l( y-1 \r)}}{dx}}\r)}
\r\}
\ee

\ni According to \cite{abel3}, it is possible to construct a transformation
mapping \eq{AH_2} into a canonical form of AIR admitting a \PFQ solution.
However, as in the BHE case, simpler 
expressions result if we transform
\eq{AH_2} into a non-canonical AIR equation. The transformation used for
this purpose is


\be{TR2_2}
x \rightarrow \fr{1}{x} - \fr{s_{{1}}}{2}-\lambda,
\ee

\ni which maps \eq{AH_2} into the Abel equation


\be{AIR_1F1_CHE}
\y1={\fr {y}{ \l( 2\,\lambda\,x - 1 \r) y + 2\,{\lambda}^{2} \l( \tau+\sigma \r) {x}^{2}-2 \l( \tau+1 \r)\lambda\, x + 1 }}
\ee

\ni Following \cite{abel3}, this AIR equation can be transformed into a \PFQ
one by combining the \yx\ transformation with transformation \eq{riccati_to_linear} mapping a Riccati into a
second order linear equation; the combination results in


\be{TR3_2}
\l\{
x \rightarrow -{\fr {x\,\y1}{ 2 \l( \tau+ \sigma \r) {\lambda}^{2}\,y}}
\ \ 
y \rightarrow x
\r\} 
\ee

\ni leading to


\be{1F1_CHE}
\yt={\fr { 2 \l( x - \tau - 1 \r) \lambda - 1 }{x}}\ \y1
+ {\fr { 2\l( \tau+\sigma \r) {\lambda}^{2} \l( x-1 \r) }{{x}^{2}}}\ y
\ee

\ni Finally, using $\{ x \rightarrow 2\,\lambda\,x,\, y \rightarrow {x}^{\l(
1+\tau-\sqrt {1-2\, \sigma+{\tau}^{2}} \r) \lambda }y \}$, equation \eq{1F1_CHE}
is obtained from the confluent \1F1 hypergeometric equation \eq{1F1_mu_nu}
\ni at $\{\mu=\l( \sigma-1+\sqrt {1-2\,\sigma+{\tau}^{2}} \r)\lambda,\,
\nu=1 + 2\,\sqrt{1-2\,\sigma+{\tau}^{2}}\,\lambda\}$, from where the solution to
\eq{1F1_CHE}, in terms of the Whittaker {\bf M}
and {\bf W} functions\footnote{${\bf M} \left( a,b,z \right) = 
{
    {{z}^{b+1/2}}\,
    {{\e^{-z/2}}}\,
    {\1F1(1/2-a+b;\,1+2\,b;\,z)}
}
$. }  \cite{abramowitz}, is

\begin{eqnarray}
\label{ans_PFQ_2}
\lefteqn{ y = {\fr {\e^{\lambda\,x}}{{x}^{\l(\tau+1 \r) \lambda+1/2}} }}
& &
\\*[0.1 in]
& &
\nonumber
\l( {\rm \bf M}\l(\fr{1}{2} + ( 1-\sigma )\, \lambda,\, \sqrt{1-2\,\sigma+{\tau}^{2}}\,\lambda,\, 2\,\lambda\,x \r) C_1
+ {\rm \bf W}\l(\fr{1}{2} + ( 1-\sigma )\, \lambda,\, \sqrt{1-2\,\sigma+{\tau}^{2}}\,\lambda,\, 2\,\lambda\,x\r) C_2
 \r)
\end{eqnarray}

Summarizing, departing from the Heun confluent equation \eq{CHE_2} we have
arrived at the \1F1 equation \eq{1F1_CHE} with solution \eq{ans_PFQ_2}
through a process of the form $ Heun \rightarrow Abel \rightarrow
Abel_{_1F_1 solvable} \rightarrow \1F1 $. The three transformations 
\eq{TR1_2}, \eq{TR2_2} and \eq{TR3_2} can be combined into one
transformation, 


\be{tr_tot_2}
y \rightarrow
{\fr {{x}^{-\l( \tau+1 \r) \lambda + 1/2}}{\sqrt {x-1}}}\ {\exp\l({\lambda\,x + 2\,{\lambda}^{2} 
\l( \tau+\sigma \r) \int \!{\fr { \l( x-1 \r) y}{{x}^{2}\y1}}\,{dx} }\r)},
\ee

\ni which maps the CHE \eq{CHE_2} into the \1F1 equation \eq{1F1_CHE} in one
step. A closed form solution for the CHE \eq{CHE_2} when $\sigma^2 \neq
\tau^2$ is then given by \eq{tr_tot_2}, where on the ``right-hand-side" the
value of $y$ is given by \eq{ans_PFQ_2}. Like \eq{tr_tot} in the BHE case,
this solution \eq{tr_tot_2} can also be expressed as a linear combination of
\1F1 functions with non-constant coefficients and free of integrals - see
\eq{tr_tot_2_free_of_integrals}. An independent check for correctness of
these solutions was also performed using symbolic computation software.


\section{Closed form solutions for a subfamily of the GHE}
\label{GHE}

Solutions for the GHE \eq{CHE_NF} are obtained from \eq{can_forms} by taking $P(y) = y\,(y-1)$, that
is, departing from


\be{AIR3}
\y1={\fr {y\,(y-1)}{ \l( s_{2}\,x^2+s_{1}\,x+s_{0} \r) y + r_{2}\,x^2 + r_{1}\,x+r_{0}}}
\ee

\ni The steps to construct these solutions are the same as those of
the previous sections. Using 

\be{inv_TR1}
\{
x \rightarrow { \fr{x\,( 1-x)\,y'}{( s_{{2}}x + r_{{2}} )\, y}},\, y \rightarrow x \}
\ee

\ni the AIR equation \eq{AIR3} is transformed into the linear ODE

\be{AL_2F1}
\yt = {\fr { \l( s_{{2}} \l( s_{{1}}-1 \r) {x}^{2}
    + \l( (s_{{1}} -2)\,r_{{2}} + s_{{2}}\, r_{{1}} \r) x 
    + r_{{2}} \l( 1 + r_{{1}} \r)  \r) }
    {x \l( s_{{2}}\, x + r_{{2}} \r)\,  \l( x-1 \r) }}\, \y1 
- {\fr { \l( s_{{0}}\,x + r_{{0}} \r)  \l( s_{{2}}\, x + r_{{2}} \r) }
    { {x}^{2}\l( x-1 \r)^{2}}}\, y
\ee

\ni This is a Heun equation of the form \eq{heun}, with four regular
singularities at $\{ 0,1,-r_2/s_2,\infty\}$. For $s_2=0$, \eq{AL_2F1}
simplifies to a Gauss equation with \2F1 solutions. When $s_2 \neq 0$, in \eq{AIR3} one
can take $s_2=1$, and, putting $r_2=-a$, the
singularities of \eq{AL_2F1} are fixed at $\{ 0,1,a,\infty\}$, resulting in a non-trivial
Heun family depending on four parameters. The relation between the
Abel parameters $\{r_i,s_i\}$ and the Heun parameters $\{A,B,D,E,F\}$ is
obtained by rewriting \eq{AL_2F1} in normal form and comparing
coefficients:

\begin{eqnarray}
\label{Heun_to_Abel_parameters}
\nonumber
A & = & r_{{0}}-(2\,r_{{0}} + s_{{0}})\, a - \fr{ 1}{2}\l( \l(r_{{1}} + 1\r) \l(s_{{1}} + \fr{1}{a}\r) + {r_{{1}}}^{2} -1 \r),
\\
\nonumber
B & = & {\fr {r_{{1}}+s_{{1}}a+1 - 2\l( A\,{a}^{2} + (1- A)\, a\r) }{2\,a \l( a-1 \r) }},
\\
D & = & \fr {1 - {r_{{1}}}^{2}}{4} - r_{{0}}\,a,
\\
\nonumber
E & = & \l( 1 - a \r) \l(  \l( A+B \r) ^{2}a- \l( A+B \r)  \l( A+B-1 \r) +{\fr {D-A}{a}}-{\fr {D}{{a}^{2}}} \r),
\\
\nonumber
F & = & -\fr{3}{4}
\end{eqnarray}

\ni So, with
respect to the most general case \eq{GHE_NF}, the restriction in the GHE \eq{AL_2F1} under consideration  consists of fixing one of the
six Heun parameters, $F$, and expressing another one, $E$, as a function of
those remaining. To derive a solution to \eq{AL_2F1}, the equation is
first written in normal form and the parameters $\{s_0,r_0,r_1\}$ are 
redefined in terms of new parameters $\{\Delta,\sigma,\tau\}$ using

\be{tau_sigma_Delta_GHE}
s_{{0}}= \fr{{s_{{1}}}^{2}}{4} -{\Delta}^{2},
\ \ \ \ 
r_{{0}}= \l( {\Delta}^{2} - \fr{{s_{{1}}}^{2}}{4} \r) a- 2\,\sigma\,\Delta+s_{{1}}\tau,
\ \ \ \ 
r_{{1}} = 2\,\tau -a\,s_{{1}}
\ee

\ni aiming at showing that there are only four independent
parameters and at separating non-Liouvillian from Liouvillian
special cases of the solution, which will happen at $\sigma = \pm \tau$. The
dependence of \eq{tau_sigma_Delta_GHE} on $\{\tau,\sigma\}$ is derived as in the previous sections, from
the canonical form of the AIR \eq{A_GHE} (see comments after
\eq{AIR_1_canonical} and also \cite{abel2}).
The new parameter $\Delta$
is related to the Heun parameter $A$ through $s_0$ and the first equation
in \eq{Heun_to_Abel_parameters}, and is introduced here to avoid square roots
in the transformation formulas\footnote{The use of $\lambda^2 = -A$ in \se{CHE}
brings the same advantage.}. With this notation, the AIR \eq{AIR3} becomes

\be{A_GHE}
\y1 = {\frac
    {y \l( y-1 \r) }
    { \l( {x}^{2} + s_{{1}}x + {s_{{1}}}^{2}/4 - {\Delta}^{2} \r) y 
    - a\,{x}^{2} + \l( 2\,\tau -a\,s_{{1}} \r) x - \l(  {s_{{1}}}^{2}/4 - {\Delta}^{2} \r) a - 2\,\sigma\,\Delta + s_{{1}}\tau}
}
\ee

\ni and the Heun equation \eq{AL_2F1} in normal form, at $\{s_2=1,r_2=-a\}$, 
appears directly expressed in terms of the four irreducible parameters
$\{a,\Delta,\sigma,\tau\}$ as


\begin{eqnarray}
\label{GHE_3}
\lefteqn{ \yt = 
\l( 
{\fr 
    {2\,{a}^{2} \l( a-1 \r) {\Delta}^{2} - 2\sigma\,a \l( 2\,a -1\r) \Delta +  \l( 2\,{\tau}^{2} -1/2\r) a + \tau + 1/2} 
    {a\,x}}
\r.
}
& &
\\*[0.1 in]
\nonumber
& &
- {\fr 
    {2\l( a \l( a-1 \r)^{2}{\Delta}^{2} - \sigma\, \l( 2\,a -1\r)  \l( a-1 \r) \Delta + \l( \tau - 1/2\r) \l( \l( \tau + 1/2 \r) a - \tau \r)\r) }
    {\l( a-1 \r) \l( x-1 \r) }}
+ {\fr 
    {\tau -a + 1/2 }
    { a\l( a-1 \r) \l( x-a \r)}}
\\*[0.1 in]
\nonumber
& &
\l.
+ {\fr 
    {{a}^{2}{\Delta}^{2} - 2\,a\,\sigma\,\Delta + {\tau}^{2} - 1/4}
    {{x}^{2}}}
+ {\fr 
    {\l( a-1 \r) ^{2}{\Delta}^{2} - 2\,\sigma\, \l( a-1 \r) \Delta + {\tau}^{2} - 1/4}
    {\l( x-1 \r) ^{2}}} 
+ \fr{3}{4\, \l( x-a \r) ^{2}}
\r) y
\end{eqnarray}

\subsection{Liouvillian solutions for the GHE \eq{GHE_3} when $\sigma = \pm \tau$}
\label{GHE_Liouvillian}

\ni By construction, as in the previous sections, for $\sigma =
\pm \tau$, the GHE \eq{GHE_3} admits Liouvillian solutions computable using Kovacic's method; for $\sigma = \tau$, 


\begin{eqnarray}
\label{Liouvillian_sol_GHE3_1}
\lefteqn{ y = {\fr 
    {{x}^{\tau - a\Delta+1/2} \l( x-1 \r)^{ \l( a-1 \r) \Delta - \tau + 1/2}}
    {\sqrt {a-x}}}}
& &
\\*[0.1 in]
& &
\nonumber
\l( 
C_1
+ \l( 
    {\bf B}_{x} \l( 1 + 2\l(a\,\Delta - \tau \r),\, 2\l( \l( 1-a \r) \Delta + \tau \r) \r)  
    -a\,{\bf B}_{x} \l( 2 \l(a\,\Delta - \tau \r),\, 2 \l( \l( 1-a \r) \Delta + \tau \r) \r) 
    \r) C_2 
\r)
\end{eqnarray}

\ni where ${\bf B}_x$ (one index and two arguments - see (6.6.1) in \cite{abramowitz}) is
the incomplete beta function. The Liouvillian form of the solution of \eq{GHE_3} at $\sigma = -\tau$
is

\begin{eqnarray}
\label{Liouvillian_sol_GHE3_2}
\lefteqn{ y = {\fr 
    {{x}^{\tau + a\Delta+1/2}
    \l( x-1 \r)^{\l( 1-a \r) \Delta-\tau+1/2}}
    {\sqrt {a-x}}}}
& &
\\*[0.1 in]
& &
\nonumber
\l( 
    C_1+ C_2 \l( 
        {\bf B}_{x} \l( 1 -2 \l( a\,\Delta+\tau \r),\, 2 \l( \l( a-1 \r) \Delta + \tau \r)  \r) 
        - a\,{\bf B}_{x} \l( -2 \l( a\,\Delta+\tau \r),\, 2 \l(  \l( a-1 \r) \Delta+\tau \r)  \r)  
    \r) 
\r)
\end{eqnarray}

\ni The existence of solutions in terms of the incomplete Beta functions is
also pointed out in \cite{isca2001}, where a solution for the GHE as a power
series is constructed, and an approximate solution involving a combination
of two incomplete Beta functions is derived - see also \cite{isca2003}.

\subsection{A solution in terms of \2F1 functions for the GHE \eq{GHE_3} when $\sigma^2 \neq \tau^2$}

As in the case of the BHE and CHE equations, a solution for the GHE
\eq{GHE_3} when $\sigma^2 \neq \tau^2$ is constructed by composing a
transformation mapping \eq{GHE_3} into the Abel AIR \eq{A_GHE} with a
transformation
mapping \eq{A_GHE} into a second order linear equation admitting
hypergeometric solutions, in this case of \2F1 type. 

The derivation of results below follows the same path shown in the previous
sections. Summarizing, the transformation mapping
the GHE \eq{GHE_3} being solved into the AIR \eq{A_GHE} is


\be{TR1}
\left\{ 
    x \rightarrow y,\,
    y \rightarrow  
    \e^{\int \!{\fr 
    {\l(  \l( x + (s_{{1}}-1)/2 \r) {y}^{2}+ \l(  \l( 1 -2\,x-s_{{1}} \r) a + \tau \r) y + \l( \l( s_{{1}}/2 + x \r) a-\tau-1/2 \r) a \r) y' }
    {y \l( y-1 \r)  \l( a-y \r) }}\,{dx}} 
\right\} 
\ee

\ni The transformation mapping the AIR \eq{A_GHE} into a \PFQ second order
linear equation is

\be{TR23}
\left\{ 
    x \rightarrow -2\,{\fr {\Delta\, \l( \tau+\sigma \r) y}{x \l( x-1 \r)\y1 }} - \fr{s_{{1}}}{2} - \Delta,
    \ \ 
    y \rightarrow x
\right\} 
\ee

\ni and the resulting equation, of \2F1 type, is


\be{2F1_GHE}
\yt = {\fr 
    {\l( 2\,\l(\Delta - 1 \r) x + 1 - 2\,(a\,\Delta + \tau) \r)}
    {x \l( x-1 \r)}}\,\y1
    + {\fr 
    {2\,\l( x-a \r) \Delta\, \l( \tau+\sigma \r)}
    {{x}^{2}\l( x-1 \r) ^{2}}}\, y
\ee

\ni Combining the transformations \eq{TR1} and \eq{TR23}, a transformation
mapping \eq{GHE_3} into \eq{2F1_GHE} in one step is

\be{tr_tot_3}
y \rightarrow 
    {\fr 
        {{x}^{a\,\Delta+\tau+1/2}\,  \l( x-1 \r)^{\l( 1-a \r) \Delta-\tau+1/2}}
        {\sqrt {a-x}}}
    \ {\exp \l({2\,\Delta\, \l( \tau+\sigma \r) \int \!{\fr {\l( x-a \r)y }{ (x-1)^{2}\, {x}^{2}\,y'}}\,{dx}}\r)}
\ee

\ni Hence, the solution to the Heun equation \eq{GHE_3} to which this
section is dedicated is given by the expression above, where, in the
``right-hand-side", the value of $y$ is given by the solution to
\eq{2F1_GHE}, that is,

\begin{eqnarray}
\label{sol_AL_pFq}
\lefteqn{ y = 
    {x}^{a\Delta+\tau-{\rm T}}\, 
    \l(x-1\r)^{(1-a)\,\Delta+\Sigma-\tau} }
& &
\\*[0.1 in]
& &
\nonumber
\l( 
    {\2F1(\Sigma+\Delta-{\rm T},\Sigma-\Delta+1-{\rm T};\,1-2\,{\rm T};\,x)}\, C_1
    + {x}^{2\,{\rm T}}{\2F1(\Sigma+\Delta+{\rm T},\Sigma-\Delta+1+{\rm T};\,1+2\,{\rm T};\,x)}\, C_2
\r)
\end{eqnarray}

\ni where, to make the structure of this solution visible, instead of $\{\sigma,\tau\}$ we are using

\be{Sigma_Tau}
\Sigma = \sqrt{{ \l( a-1\r) ^{2}{\Delta}^{2}-2\l( a-1 \r)\sigma\,\Delta + {\tau}^{2}}},
\ \ \ 
{\rm T}= \sqrt{{{a}^{2}{\Delta}^{2} -2\,a\,\sigma\, \Delta + {\tau}^{2}}}
\ee

\ni This solution \eq{sol_AL_pFq} in turn is computed noting that \eq{2F1_GHE} is
obtained by changing variables 

\be{tr_AL_pFq}
y \rightarrow {{ {x}^{{\rm T}-a\,\Delta-\tau}
\l( x-1 \r)^{\tau -\Sigma + \l( a-1 \r) \Delta}}\, y}
\ee

\ni in Gauss' \2F1 equation

\begin{equation}
\l( {x}^{2}-x \r) \yt + \l(  \l( \mu+\nu+1 \r) x-\rho \r) \y1 + \mu\,\nu\,y = 0
\ee

\ni taken at $\{\mu=\Sigma+\Delta-{\rm T},\, \nu=\Sigma-\Delta+1-{\rm T},\,
\rho=1-2\,{\rm T}\}$. Hence, the same transformation \eq{tr_AL_pFq} maps the
solution of Gauss' equation into \eq{sol_AL_pFq}. For a solution equivalent
to \eq{tr_tot_3}, free of integrals, expressed as a linear combination of \2F1
functions with non-constant coefficients, see
\eq{tr_tot_3_free_of_integrals}. These and the Liouvillian solutions
presented for the GHE \eq{GHE_3} were also verified for correctness using
symbolic computation software.

\section{Alternative derivation of solutions free of integrals}
\label{alternative}

In the previous sections, non-Liouvillian solutions, as well as their 
Liouvillian special cases and the relationship between the Heun parameters for their existence, were
derived for the BHE, CHE and GHE equations \eq{BHE_1}, \eq{CHE_2} and
\eq{GHE_3}. The non-Liouvillian solutions \eq{tr_tot}, \eq{tr_tot_2} and
\eq{tr_tot_3}, however, have the drawback of containing non-trivial
uncomputed integrals. In this section, from the knowledge of the form of
these solutions and exploring non-local transformations,
equivalent solutions free of integrals are derived. 

We start by
recalling that second order linear equations

\be{generic_linear_ode}
\yt = c_1\, \y1 + c_0\, y 
\ee

\ni where $c_i \equiv c_i(x)$, can always be mapped into Riccati equations
\eq{riccati} back and forth. The transformation mapping
\eq{generic_linear_ode} into a Riccati equation is of the form

\be{linear_to_riccati}
y \rightarrow {\e^{-\int G(x)\,y\, {dx}}}
\ee

\ni where $G(x)$ is an arbitrary function, and the transformation mapping a
Riccati equation \eq{riccati} into a linear equation is given by \eq{riccati_to_linear}. It is also known
that any two Riccati equations can be mapped between themselves through
M\"obius transformations of the dependent variable $y$ with variable
coefficients $f_i \equiv f_i(x)$,

\be{MobiusY}
y \rightarrow \fr {f_1\, y + f_2} {f_3\, y + f_4},
\ee

\ni where $f_{{1}} f_{{4}} -f_{{3}} f_{{2}} \neq 0$. If we now transform
\eq{generic_linear_ode} into a Riccati equation using \eq{linear_to_riccati},
then apply the M\"obius transformation \eq{MobiusY}, and to the resulting
equation we apply transformation \eq{riccati_to_linear}, we obtain
another second order linear equation. 
The composition of these three transformations is the non-local transformation

\be{MobiusR}
y \rightarrow {\exp{\l(-\int \!{\fr {f_{{1}} \Omega\,y' + f_{{2}}
\Upsilon\, y}{f_{{3}} \Omega\,y' + f_{{4}} \Upsilon\,y}}\,{dx}\r)}},
\ee

\ni where

\begin{equation}
\Omega=f_{{1}} f_{{4}}  -f_{{3}} f_{{2}} \neq 0, 
\ \ \ \ 
\Upsilon = c_{{0}}\,  f_{{3}}^{\,\,2} + \left( {f_{{1}}}{'}-c_{{1}} f_{{1}}
\right) f_{{3}} - \left( {f_{{3}}}{'}+f_{{1}} \right) f_{{1}} \neq 0,
\ee

\ni and, in fact, this transformation suffices to generate the whole
class of linear equations from any given one. For example, applying
\eq{MobiusR} at $f_1 = f_4 = 1$ to $\yt = 0$, we obtain an equation as
general as \eq{generic_linear_ode}. 

The particular case of \eq{MobiusR} at $f_1 = f_4 = 0$ and $f_2 = f_3 =1$,

\be{MobiusR_0110}
y \rightarrow {\exp{\l(\int \!{\fr {c_{{0}}\,y}{\y1}}{dx}\r)}}
\ee

\ni is relevant to the results of the previous sections: the 
transformations \eq{tr_tot}, \eq{tr_tot_2} and \eq{tr_tot_3},
respectively mapping the BHE, CHE and GHE into \PFQ equations, are in fact compositions of
transformations of the form \eq{MobiusR_0110} with transformations
$y \rightarrow P(x)\,y$. 
The linear equation obtained by applying \eq{MobiusR_0110} to
\eq{generic_linear_ode} is

\be{generic_linear_ode_R}
\yt= \l( {\fr { {c_{{0}}}{'} }{c_{{0}}}} -c_{{1}}\r) \y1 + c_{{0}}\,y
\ee

\ni By applying to this equation the same transformation \eq{MobiusR_0110},
we reobtain\footnote{The composition of \eq{MobiusR_0110} with itself is
equal to the identity transformation.} \eq{generic_linear_ode}. 
So, $y$ in the ``left-hand-side" of \eq{MobiusR_0110} represents
the solution to \eq{generic_linear_ode_R} or \eq{generic_linear_ode}, respectively written in
terms of the solution to \eq{generic_linear_ode} or
\eq{generic_linear_ode_R}, represented by $y$ in the ``right-
hand-side" of \eq{MobiusR_0110}. As we shall see in the following subsections, this is indeed the mechanism by which 
solutions to Heun equations (here represented by \eq{generic_linear_ode_R}) were expressed as 
exponentials of integrals of solutions to \PFQ equations (here represented by
\eq{generic_linear_ode}) in the previous sections.


%
%

With this understanding of matters, however, it is possible to show that, despite the integral sign entering
\eq{MobiusR_0110}, the solution to \eq{generic_linear_ode_R} can be derived
without performing any integration. For that purpose,
we note that, given a generic
linear ODE in $y$, it is always possible to construct the generic linear ODE of the
same order satisfied by $p = \y1$. Concretely, if $y$ satisfies
\eq{generic_linear_ode}, the ODE for $p$ is\footnote{In the easier case
$c_0=0$, the equation for $p$ is: ${\it p''}=c_{{1}}{ p'}+{c_{{1}}}{'}p$.}

\be{ode_y1}
p''
= 
\l({\fr { {c_{{0}}}{'}}{c_{{0}}  }} + c_{{1}} \r) {\it p'} 
+ \l({c_{{1}}}{'} + c_{{0}} - {\fr { {c_{{0}}}{'}c_{{1}}  }{c_{{0}}  }}\r) p
\ee

\ni and by substituting $y = \int
\! p\ dx$ into \eq{generic_linear_ode}, we obtain

\begin{equation}
y = \int \! p\ dx = \fr{p' - c_1\,p}{c_0}
\ee

\ni Therefore, when \eq{ode_y1} can be solved, the solution $y$ for
\eq{generic_linear_ode} can be obtained directly from $p$ by differentiation. For
instance, writing $p$ in terms of some $f \equiv
f(x)$ and $g \equiv g(x)$, as

\begin{equation}
 p = f\,C_1 + g\,C_2,
\ee

\ni a solution for \eq{generic_linear_ode} is computed from $p$, without using integration, as 

\be{y_from_p}
y = 
{\fr {f\,' - c_{{1}}\, f}{c_{{0}}  }}\, C_1 
+ {\fr {g\,' - c_{{1}}\, g}{c_{{0}}  }}\, C_2
\ee

\ni Composing now the non-local transformation \eq{MobiusR_0110} with the
introduction of $p=y'$, that is, plugging the coefficients
of \eq{generic_linear_ode_R} into \eq{ode_y1}, we obtain

\be{MobiusR_p}
p''={{ \l( \fr{ 2\,{c_{{0}}}{'}}{c_0} -c_{{1}} \r) p'}}
+ \l(  c_{{0}} - {c_{{1}}}' 
    + \fr { c_{{1}}  {c_{{0}}}{'} + {c_{{0}}}{{ ''}} } {c_{{0}}}  
    - 2 \l( \fr {{c_{{0}}}{'}}{c_0} \r)^{2} \r) p,
\ee

\ni and the key observation is that the normal form\footnote{For
rewriting equations in normal form, see the Appendix.} of \eq{MobiusR_p}
{ is the same as that of} \eq{generic_linear_ode}. Consequently, if
\eq{generic_linear_ode} is of \PFQ type\footnote{An equation is
of \PFQ type if it admits solutions of the form \eq{hyper3_sol}; these solutions can be 
computed systematically \cite{hyper3}.}, then \eq{MobiusR_p} is too, even when
\eq{generic_linear_ode_R} may not be (in what follows it will be of Heun
type), and the solution to \eq{generic_linear_ode_R} can be expressed not
just using the integral form
\eq{MobiusR_0110}, but also using \eq{y_from_p} as a linear combination, with
variable coefficients, of the \PFQ solutions $\{f,g\}$ of \eq{MobiusR_p} and their derivatives.


Based on these observations, the derivation of the solutions
\eq{tr_tot}, \eq{tr_tot_2} and \eq{tr_tot_3} for the BHE, CHE and GHE
equations \eq{BHE_1}, \eq{CHE_2} and \eq{GHE_3}, performed in the previous
sections, can be reformulated entirely, shortcutting the Abel equation step,
and resulting in solutions free of integrals as follows. 

\subsection{Solution free of integrals for the BHE \eq{BHE_1}}
\label{BHE_alternative}

Applying transformation \eq{MobiusR_0110} to the \1F1
equation \eq{1F1_BHE}, derived from the AIR
\eq{AIR_BHE_alternative} in \se{BHE}, we obtain


\be{BHE_intermediary}
\yt= \l( 2 \l(\tau - x \r) + \fr{1}{x} \right) \y1 + 2 \l( \tau + \sigma \right) x\, y
\ee

\ni This equation is already of Heun type, and by rewriting it in normal form\footnote{ For the formula to
write linear equations in normal form, see the Appendix.}, that is, changing
further

\be{P_BHE}
y \rightarrow {\exp{\l(\int \!\fr{1}{2\,x} - x + \tau\, {dx}\r)}}\, y
\ee

\ni we directly obtain the two parameter BHE \eq{BHE_1}, derived from the
AIR \eq{AIR1}. Hence, combining these two transformations \eq{MobiusR_0110}
and \eq{P_BHE}, we also directly obtain the solution \eq{tr_tot} computed in
\se{BHE} for the BHE \eq{BHE_1}.

We note here that the particular form \eq{1F1_BHE} of the \1F1 equation
derived in \se{BHE}, which has only one irregular singularity at infinity,
has the important feature that under the transformation \eq{MobiusR_0110},
\eq{1F1_BHE} gains one regular singularity at the origin. Hence, the
resulting equation \eq{BHE_intermediary} is not a \1F1 equation anymore but
a 2-parameter biconfluent Heun equation. Both the augmentation in the number
of singularities under transformation \eq{MobiusR_0110} and the change
in type from \1F1 to BHE do not happen with all \1F1 equations. 

Now, since this BHE equation \eq{BHE_intermediary} was obtained from a \1F1
equation using the transformation \eq{MobiusR_0110}, as explained, the
derivative $p \equiv \y1$ of the solution to \eq{BHE_intermediary} also
satisfies a \1F1 equation. According to \eq{ode_y1}, the equation for $p =
y'$ associated to \eq{BHE_intermediary} is

\be{BHE_ode_y1}
{\it p''}={\fr { 2 \l( 1 + \tau\,x - {x}^{2} \r) }{x}}\, {\it p'}
- {\fr { 2 \l( 1 + \tau \,x - \l( \tau + \sigma \r) {x}^{3} \r)}{{x}^{2}}}\, p
\ee

\ni The solution to this \1F1 equation can be expressed in terms of the
Kummer functions M and U \cite{abramowitz} as


\begin{equation}
p = x\, {\e^{x \l( \tau - \sigma-x \r) }} 
\l(
{{\rm M}\l( \fr{ {\tau}^{2} -{\sigma}^{2}}{4}, \,\fr{1}{2},\, \l( x+\sigma \r) ^{2}\r)}\, C_1 
+ {{\rm U}\l( \fr{ {\tau}^{2} -{\sigma}^{2}}{4}, \,\fr{1}{2},\, \l( x+\sigma \r) ^{2}\r)}\, C_2 
\r) 
\ee

\ni Substituting this solution into \eq{y_from_p}, we obtain the solution to
the BHE \eq{BHE_intermediary}, and further applying the transformation \eq{P_BHE}, we
directly obtain the general solution, free of integrals, for the BHE
\eq{BHE_1} of \se{BHE}, as


\begin{eqnarray}
\nonumber
\lefteqn{
    y = \fr {\e^{-\sigma\,x-{x}^{2}/2}} {\sqrt {x} \l( x+\sigma \r) }
\bigl
    \l(
        \Lambda \l( x \r)\, {{\rm U}\l(\fr{{\tau}^{2} - {\sigma}^{2}}{4},\, \fr{1}{2},\, \l( x+\sigma \r) ^{2}\r)} 
        -4\,{{\rm U}\l(\fr{{\tau}^{2} - {\sigma}^{2}}{4} - 1,\, \fr{1}{2},\, \l( x+\sigma \r) ^{2}\r)}
    \r) C_1
\right.
}
\\*[.1in]
\label{tr_tot_1_free_of_integrals}
 & + &
\left.
    \l(
        \l( {\tau}^{2}-{\sigma}^{2}-2 \r) {{\rm M}\l(\fr{{\tau}^{2} - {\sigma}^{2}}{4}-1,\, \fr{1}{2},\, \l( x+\sigma \r) ^{2}\r)}
        -\Lambda \l( x \r)\, {{\rm M}\l( \fr{{\tau}^{2} - {\sigma}^{2}}{4},\fr{1}{2}, \l( x+\sigma \r) ^{2}\r)} 
    \r) C_2
\bigr
\end{eqnarray}

\ni where $ \Lambda \l( x \r) \equiv {\sigma}^{2}+{\tau}^{2} + 2\,(2\,{x}^{2} +
\l( 3\,\sigma - \tau \r) x - \sigma\,\tau-1) $ and $\sigma^2 \neq \tau^2$.
Symbolic computation input for verifying this solution is found in the
Appendix.

This approach to the solution of the BHE \eq{BHE_1} is straightforward,
clearly simpler than the calculations presented in \se{BHE}. As shown in the
following two subsections, this simpler approach also leads to solutions
free of integrals for the CHE \eq{CHE_2} and the GHE \eq{GHE_3} discussed in
\se{CHE} and \se{GHE}. We note, however, that without the knowledge of the
equations of Heun and \PFQ type being linked, or of the form of the
transformation relating them\footnote{In the BHE case, these are \eq{BHE_1},
\eq{1F1_BHE} and \eq{tr_tot}.}, both of which were derived in the previous sections from the connection
$\mbox{Heun} \leftrightarrow \mbox{Abel}$, the existence of the straightforward mechanism
used in this section is not evident. 


\subsection{Solution free of integrals for the CHE \eq{CHE_2}}
\label{CHE_alternative}

Like in the BHE case, a solution to the CHE \eq{CHE_2}, equivalent
to \eq{tr_tot_2} and free of integrals, can be obtained by first applying
\eq{MobiusR_0110} to the \1F1 equation \eq{1F1_CHE}, derived from the AIR
\eq{AIR_1F1_CHE} in \se{CHE}, leading to a CHE,


\be{CHE_intermediary}
\yt=
{\fr { 1 + 2\,\lambda \l( x-1 \r)  \l( 1-\tau -x \r)  }{x \l( x-1 \r) }}\,\y1 
+ {\fr {2\,{\lambda}^{2} \l( x-1 \r)  \l( \tau+\sigma \r)}{{x}^{2}}}\, y 
\ee

\ni Rewriting this equation in normal form directly results in the CHE
\eq{CHE_2} discussed in \se{CHE}. So, the problem now is the computation of
solutions to the CHE \eq{CHE_intermediary}. As in the BHE case,
we know, by construction, that when $y$ satisfies \eq{CHE_intermediary}, $p \equiv y'$
satisfies a \1F1 equation. According to \eq{ode_y1}, the equation for
$p$ is


\begin{eqnarray}
\lefteqn{
p'' = 
{\fr { 3 -2\,\lambda \l( 1+\tau \r) + \l( 2\,\lambda\l( \tau+2 \r)  -1 \r) x  -2\,\lambda\,{x}^{2} }{x \l( x-1 \r) }}\, p' 
} & & 
\\ 
\nonumber
& + & {\fr { 2\,{\lambda}^{2} \l( x-1 \r) ^{3} \l( \tau+\sigma \r) -2\, \l( x-1 \r)  \l( {x}^{2}-2\,x + \tau+1 \r) \lambda-1-x} {{x}^{2} \l( x-1 \r) ^{2}}}\,p 
\end{eqnarray}

\ni The general solution to this equation can be written in terms of Whittaker functions {\bf M} and {\bf W}
\cite{abramowitz} as


\be{sol_p_CHE}
p  =  \fr {\l( x-1 \r) {x}^{\l( \tau+1 \r) \lambda -3/2} } {\e^{\lambda\,x}} 
   \l( {\bf M} \l(\mu,\, \nu,\, 2\,\lambda\,x \r)\, C_1  + {\bf W} \l(\mu,\, \nu,\, 2\,\lambda\,x \r)\, C_2 \r)  
\ee


\ni where $\mu= \lambda\l( 1-\sigma \r) + 1/2$ and
$\nu=\lambda\,\sqrt {{\tau}^{2}-2\,\sigma+1}$. Substituting this solution
into \eq{y_from_p} leads to a solution free of integrals for the CHE
\eq{CHE_intermediary}, from where the solution to its normal form
\eq{CHE_2} of \se{CHE}, is


\begin{eqnarray}
\label{tr_tot_2_free_of_integrals}
\nonumber
\lefteqn{ y = {\frac {1}{\sqrt {x-1}}}
\bigl
    \l(
        {{\rm \bf W}\l(\mu,\,\nu,\,2\,\lambda\,x\r)} 
        + \lambda\, \l( \tau-\sigma \r) {{\rm \bf W}\l(\mu-1,\,\nu,\,2\,\lambda\,x\r)}
     \r) C_2
\right.
} & & 
\\
& & 
\left.
    + \l(
        \lambda \l( \tau+\sigma \r) {{\rm \bf M}\l(\mu,\,\nu,\,2\,\lambda\,x\r)}
        + \l(  \l( 1-\sigma \r) \lambda-\nu \r) {{\rm \bf M}\l(\mu-1,\,\nu,\,2\,\lambda\,x\r)}
     \r) C_1
\bigr
\end{eqnarray}

\ni Symbolic computation input for verifying this solution is found in the
Appendix.

\subsection{Solution free of integrals for the GHE \eq{GHE_3}}
\label{GHE_alternative}

The derivation done in \se{GHE} of a solution for the GHE \eq{GHE_3} can be
reformulated as in the BHE and CHE cases. Applying \eq{MobiusR_0110} to the
\2F1 hypergeometric equation \eq{2F1_GHE}, we obtain


\be{GHE_intermediary}
\yt = 
    {\fr 
        { \l( 2\,\Delta+1 \r) {x}^{2} - 2\l(  \l( 2\,\Delta+1 \r) a+\tau \r) x+2\,{a}^{2}\Delta+ \l( 2\,\tau+1 \r) a }
        {x \l( x-1 \r)  \l( a-x \r) }}\, \y1 
    - {\fr 
        { 2 \l( a-x \r) \Delta \l( \tau+\sigma \r) y}
        {{x}^{2} \l( x-1 \r) ^{2}}}
\ee

\ni which is a GHE with four regular singularities at $\{ 0, 1,a,\infty \}$. Rewriting \eq{GHE_intermediary} in normal form  using 

\be{P_GHE}
y \rightarrow {\exp{\l(\int \!
    {\frac 
        {2\, \l( x-a\r) \tau -2\, \l( a-x \r)^{2} \Delta + a \l( 2\,x-1 \r) -{x}^{2}}
        {2\,x \l( x-1 \r) \l( x-a \r) }}
    \,{dx}\r)}
}\, y
\ee

\ni we obtain the 4-parameter GHE \eq{GHE_3} of \se{GHE}. The solution
\eq{tr_tot_3} presented in \se{GHE} for this equation is identical to the
composition of these two transformations \eq{MobiusR_0110} and \eq{P_GHE}.
As in the previous subsections, using \eq{ode_y1} we compute the equation
for $p = \y1$ associated to \eq{GHE_intermediary}, which, as discussed, is by
construction a \2F1 equation when written in normal form. Solving for $p$, from
\eq{y_from_p}, we obtain the solution to \eq{GHE_intermediary}, and applying
\eq{P_GHE} to it, we obtain a solution free of uncomputed integrals for the
GHE \eq{GHE_3},


\begin{eqnarray}
\label{tr_tot_3_free_of_integrals}
\lefteqn{ y = {\frac {\l( x-1 \r) ^{\Sigma+1/2}}{\sqrt {x-a}}}
    \bigl
    \r.
    }
& & 
\\
\nonumber
& &
\!\!\!\!\!
    C_1 \l(
       { \fr {( {\rm T}-\Sigma-\Delta )  ( {\rm T}-\Sigma+\Delta-1 ) ( {x}^{5/2-{\rm T}}-{x}^{3/2-{\rm T}})}{2} } 
            \, {\2F1(\Sigma+\Delta-{\rm T}+1,\Sigma-\Delta-{\rm T}+2;\,2\,(1-{\rm T});\,x)}
    \r.
\\
\nonumber
& &
\!\!\!\!\!
   + \l.
       \l( {\rm T}-\fr{1}{2} \r) \l(\l( a\Delta-{\rm T}+\tau \r){x}^{1/2-{\rm T}} + \l( {\rm T}-\Sigma-\Delta \r) {x}^{3/2-{\rm T}}  \r)
            {\2F1(\Sigma+\Delta-{\rm T},\Sigma-\Delta-{\rm T}+1;\,1-2\,{\rm T};\,x)} 
    \r)
\\
\nonumber
& &
\!\!\!\!\!
    +\, C_2 \l(
        {\fr {( {\rm T}+\Sigma+\Delta )  ( {\rm T} + \Sigma-\Delta+1 )  ( {x}^{3/2+{\rm T}} -{x}^{5/2+{\rm T}} )}{2}}
            \, {\2F1(\Sigma+\Delta+{\rm T}+1,\Sigma-\Delta+{\rm T}+2;\,2\,(1+{\rm T});\,x)}
            \right.
\\
\nonumber
& &
\!\!\!\!\!
        +
        \left.
        \l( {\rm T} + \fr{1}{2} \r) \! \l(\l( a\Delta+{\rm T}+\tau \r){x}^{1/2+{\rm T}} -  \l( {\rm T}+\Sigma+\Delta \r) {x}^{3/2+{\rm T}}  \r) 
        {\2F1(\Sigma+\Delta+{\rm T},\Sigma-\Delta+1+{\rm T};\,1+2\,{\rm T};\,x)} \r)
\l.
\!\!\!\bigr
\end{eqnarray}

\ni where $\Sigma$ and ${\rm T}$ are defined in \eq{Sigma_Tau}. This
solution is also valid when the GHE \eq{GHE_3} admits Liouvillian solutions,
that is, when $\sigma^2 = \tau^2$, although in this case the solutions
\eq{Liouvillian_sol_GHE3_1} and \eq{Liouvillian_sol_GHE3_2} are expressed in
simpler manner. Symbolic
computation input for verifying this solution \eq{tr_tot_3_free_of_integrals} is found in the Appendix.


\section{Comparison with solutions existing in the literature}
\label{comparison}

A search in the literature didn't show previous references to a link
between Heun and Abel equations as the one presented in \se{Theory}, nor a derivation of solutions to the former
equations and confluent cases from the knowledge of solutions to the latter. It is nonetheless interesting to compare the solutions for Heun equations derived through this link $Heun \leftrightarrow Abel$ and in \se{alternative}
with the ones previously presented in the literature. 
For practical reasons, the discussion is restricted to three more recent papers, by
Ronveax \cite{ronveaux2}, by Ishkhanyan and Suominen \cite{isca2003}, and by Shanin and Craster \cite{craster2002},
which present sufficiently explicit solutions for the GHE \eq{heun},
similar to the non-Liouvillian solutions
and the special Liouvillian cases derived in \se{GHE_Liouvillian} and \se{GHE_alternative}.

\subsection{Liouvillian solutions}

In \cite{ronveaux2}, the factorization of Heun's General equation \eq{heun} into a form

\be{factorization_form}
\l(L(x) {\rm D} + M(x)\r) \l(\bar{L}(x) {\rm D} + \bar{M}(x)\r) y = 0
\ee

\ni where $D \equiv d/dx$ and $\{L, M, \bar{L}, \bar{M}\}$ are polynomials,
is discussed, and six sets of conditions on the Heun parameters, such that
this type of factorization is possible, are derived, all leading to
solutions of the form

\be{solutions_via_factorization}
y = x^{\rho_1} \l(x-1\r)^{\rho_2} \l(x-a\r)^{\rho_2}
\ee

\ni for some $\rho_i$. Since these solutions are Liouvillian, they can
be computed systematically, e.g., in a symbolic computation environment like Maple or Mathematica,
where Kovacic's algorithm is implemented. 


The Liouvillian solutions \eq{Liouvillian_sol_GHE3_1} and \eq{Liouvillian_sol_GHE3_2}, 
derived here from the condition $\sigma^2 = \tau^2$ related to the canonical form
of Abel equations, are also of the form \eq{solutions_via_factorization}.
Nonetheless, the conditions for the existence of a factorization of the form \eq{factorization_form} obtained in \cite{ronveaux2}
are less general than the conditions for the existence of Liouvillian solutions derived here. For example, if in \eq{GHE_3} we change
variables $y \rightarrow \exp(y)$, the resulting Heun equation will
continue having rational coefficients and the condition $\sigma^2 = \tau^2$
will continue assuring that the solution admits Liouvillian form, computable
using Kovacic's algorithm, even when it won't be of the form
\eq{solutions_via_factorization} anymore. On the other hand, if we perform the same
change of variables in the equations obtained in \cite{ronveaux2}, the
resulting equations will be out of reach of the factorization there presented,
because $M(x)$ in \eq{factorization_form} will have an exponential factor,
while that method applies only to polynomial forms of $M(x)$.

%
%

We note that, by changing variables appropriately, the Liouvillian solutions derived in \se{Liouvillian_CHE}
and \ref{GHE_Liouvillian} for the BHE \eq{BHE_1} and CHE \eq{CHE_2}, can also be transformed 
into the form \eq{solutions_via_factorization}, which indicates
that a factorization like the one discussed in \cite{ronveaux2} exists also
for the BHE and CHE. 

\subsection{Non-Liouvillian solutions}

Non-Liouvillian solutions cannot be computed with Kovacic's algorithm, nor
is there such a general algorithm for computing them. This type of
solutions was presented in the previous sections and is discussed in other
papers in the literature. 

In \cite{isca2003}, an approach restricted to the GHE\footnote{The BHE, CHE
or other confluent cases are not discussed in \cite{isca2003}.} \eq{heun} is
discussed. Concretely, after some manipulations, the equation satisfied by $H'$, where $H$ is a
solution to the Heun equation \eq{heun}, is presented. This equation for $H'$ can be computed using
\eq{ode_y1}, and is a Fuchsian equation with five regular singularites,
located at $\{0,1,\infty ,a,q/(\alpha\,\beta) \}$. So, when
$q/(\alpha\,\beta)$ is equal to $0$, $1$ or $a$, or approaches $\infty$, the
equation has four singularities and hence both $H$ and $H'$ satisfy a Heun
equation. This happens when either $q=0$, $q = \alpha\,\beta$, $q =
a\,\alpha\,\beta$, or $\alpha\,\beta = 0$. These four
cases are presented in \cite{isca2003}, and at first sight, the approach could be compared with
the one presented here, in \se{alternative}, where Heun and related
confluent equations with the property that $H'$ is of the form \eq{hyper3_sol}, involving \PFQ functions, were derived.

The main difference between the presentation in \se{alternative} and that in
\cite{isca2003} is that, in \se{alternative}, the non-local transformation
\eq{MobiusR_0110} directly leads to non-trivial Heun equations in $H$, such
that $H'$ can be computed {\it systematically}, because it is of the form
\eq{hyper3_sol}, and from there we can systematically compute $H$, using
\eq{y_from_p}. 

On the other hand, in \cite{isca2003}, the condition that $H'$ satisfies a
Heun equation of one parameter less is not sufficient to compute its value.
So, to obtain solutions using this approach, the authors introduce additional
restrictions on the values of the Heun parameters so that $H'$ admits
solutions expressible using \2F1 functions.
Neither the origin of these ad-hoc restrictions nor a systematic manner of computing
them is shown.

For the first case, $q=0$, the additional restrictions suggested in
\cite{isca2003} are $\epsilon = -1$ and $q' = a\,\alpha' \beta'$, where
$\{q',\alpha',\beta'\}$ are some functions of the Heun parameters
$\{\alpha,\beta,\gamma,\delta,\epsilon,q\}$, together leading to

\be{isca_first_case}
H' = x\ \2F1(\alpha',\beta';\, \gamma+2;\,x)
\ee

\ni This is a case depending on only three parameters $\{\alpha',\beta',
\gamma\}$, which happens to be a particular case of the 4-parameter
GHE solved here, in \se{GHE_alternative}. That can be seen by using
\eq{ode_y1} to construct the equation for $p \equiv H'$ associated to
Heun equation \eq{GHE_intermediary} of \se{GHE_alternative}, and by noting
that its solutions are of the form \eq{hyper3_sol}

\begin{equation}
p = H' = x^{\rho_1} (x-1)^{\rho_2} (x-a)^{\rho_3}\ \2F1({\tilde \alpha},{\tilde \beta};\, {\tilde \gamma};\,x)
\ee

\ni for some $\rho_i$; that is, they depend on four parameters $\{a, {\tilde \alpha},{\tilde
\beta},{\tilde \gamma}\}$, not three. 

The other case explicitly discussed in \cite{isca2003} is $q =
\alpha\,\beta$, so a 5-parameter GHE for which a solution is shown in terms
of an infinite series of Appel functions\footnote{The Appel function is a
formal extension of the \2F1 function to two variables, expressed as a
double infinite series.} \cite{WW} which, in one case, is shown to be
expressible as an infinite sum of \2F1 functions, presented in \cite{isca2003} with
number (36). It is not evident how to
compare this formal infinite series solution with the finite-number-of-terms
solutions presented here, in \se{GHE_alternative}, nor is it evident how to
impose an additional restriction to the 5-parameter GHE treated in
\cite{isca2003} such that the formal infinite series expansion terminates. Solutions to the GHE in terms of infinite series of
\2F1 functions are also known in the literature \cite{ronveaux}, although
the authors of \cite{isca2003} make the point that the \2F1 functions
entering the formal series they present have a different
behavior than those shown in \cite{ronveaux}.

\subsubsection*{An approach based on removing ``false" singularities}

A referee has also pointed to \cite{craster2002}, a very interesting paper
by Shanin and Craster, published in 2002 but actually submitted during 2000,
where a more thorough approach to solving GHE and confluent CHE
equations (actually, the approach is for all linear equations with ``false" singularities), is presented.

The first idea in \cite{craster2002}
consists of determining relations (constraints) between the Heun
parameters such that one of the regular singularities is ``false". That will
happen when the corresponding characteristic exponents differ by an integer
but, also, no logarithmic term appears in the local expansion of the
solution. Although such an approach is entirely different from the one
developed here, where the solvable Heun equations are derived from the single Abel
AIR \eq{AIR_rho}, it is remarkable that the BHE \eq{BHE_1}, CHE
\eq{CHE_2} and GHE \eq{GHE_3} resulting from this link $Heun \leftrightarrow
Abel$ do have one such false singularity. 

In \cite{craster2002}, there is no explicit discussion of the BHE
\eq{BHE_NF}, but the approach seems applicable to that case too, and, as
is the case here, the approach does not seem to be applicable to the DHE
\eq{DHE_NF} or the THE \eq{THE_NF}. The derivation of GHE and CHE solvable
equations in \cite{craster2002} is systematic but not as straightforward
as the one-step derivation shown here in \se{CHE_alternative} and
\se{GHE_alternative}, exploring non-local transformations. Concretely, in
\cite{craster2002}, 
determining the value of the accessory parameter $q$ in \eq{heun}, such that
the equation has a false singularity,
requires using \PFQ identities and solving recurrence relations for the coefficients of series
expansions, to assure there is no logarithmic term in the solution.

The second idea presented in \cite{craster2002} is that, when the Heun
equation has a false singularity, its solution can be expressed as a linear
combination (with finite number of terms and constant coefficients) of \PFQ
functions, determined by exploring isomonodromy mappings. Although
finding the appropriate isomonodromy is a powerful idea, and the
approach is systematic, quoting the authors of \cite{craster2002} (p.628):
{\it ``The procedure of finding an appropriate isomonodromy mapping described
is quite complicated. In the relatively simple examples that we construct we
can pose an ansatz for the form of the mapping to within several unknown
parameters; these are found by direct substitution."}. The actual
procedure to determine these unknown parameters is systematic, but not so
straightforward; quoting the authors (p.630): {\it ``The simplest way to do
this is to substitute the linear combination, say $U +C\, V$ directly into the
Heun equation and then use known recursion formulae for hypergeometric
functions and their derivatives. Tedious calculations show that ... "}.

In contrast, the approach presented here, in \se{alternative}, directly
leads, by construction, to the three multiparameter BHE \eq{BHE_intermediary}, CHE \eq{CHE_intermediary} and GHE \eq{GHE_intermediary}
equations having false singularities as well as to the exact form of the linear
combinations (with non-constant coefficients) of \PFQ functions that solve these equations.

Furthermore, here, in
\se{alternative}, the use of non-local transformations links \PFQ equations
also to other linear equations, with more singularities than those of the Heun families, where again the linear combinations \eq{y_from_p} solving all
these equations involve non-constant coefficients. Although these cases can
in principle be treated by finding
isomonodromies, that approach may result non-practical as soon as the number
of singularities or the number of parameters involved increases. For
example, in perhaps the simplest case, departing from


\be{0F1_example}
{\frac {{\0F1(\ ;\,a;\,x)}}{x-\kappa}}
\ee

\ni the \PFQ equation satisfied by this expression is


\begin{equation}
\yt= \l( -{\frac {a}{x}}+ \fr{2}{ \kappa-x} \r) \y1+{\frac { \l( x-\kappa-a \r) y}{x \l( x-\kappa \r) }}
\ee

\ni Applying now the non-local transformation \eq{MobiusR_0110} we obtain


\be{example_1}
\yt= \l( {\frac {a-1}{x}} + \fr{1}{x-\kappa} + \fr {1}{x-a-\kappa} \r) \y1 + {\frac { \l( x-\kappa-a \r) y}{x \l( x-\kappa \r) }}
\ee

\ni This equation has three regular singularities at $\{0,\kappa,a+\kappa\}$
and one irregular singularity at $\infty$; they are all irreducible, and therefore
\eq{example_1} does not fit into any of the five Heun classes represented by
(\ref{GHE_NF}-\ref{THE_NF}) (it belongs to an ``upper" class). According to \se{alternative}, by
construction, the equation satisfied by $p=y'$ admits systematically computable \cite{hyper3} solutions of the form
\eq{hyper3_sol} which, when plugged
into \eq{y_from_p}, lead to the following solution to \eq{example_1}:


\begin{eqnarray}
\lefteqn{
    y = C_1\,{x}^{a} \l( a\,\, {\0F1(\ ;\,a;\,x)}- \l( x-\kappa \r) {\0F1(\ ;\,a+1;\,x)} \r)
}
& & 
\\*[0.08in]
\nonumber
& & 
+\, C_2 \l(  \l( a-2 \r)  \l(  \l( 1-a \r) \kappa + a\,x \r) {\0F1(\ ;\,2-a;\,x)}+x \l( x-\kappa \r) {\0F1(\ ;\,3-a;\,x)} \r) 
\end{eqnarray}

\ni This approach, as described in \se{alternative}, works just as straightforwardly
when we start with \PFQ functions more general than \eq{0F1_example}, while
through that process the 
equation resulting from applying \eq{MobiusR_0110} can be made to depend
on more parameters and have more singularities. Even so, by construction,
the exact linear combination (with non-constant
coefficients) of \PFQ functions solving the resulting equation is always given by
\eq{y_from_p}. Contrasting with that, depending on the starting \PFQ expression to be used in place of \eq{0F1_example}, the construction of the same
solvable cases and computation of their solutions using the approach presented
in \cite{craster2002} can be really complicated.

On the other hand, an important generalization presented in
\cite{craster2002} is that it provides a recipe for computing the
isomonodromies and constructing the related Heun equations having as solutions linear
combinations of \PFQ functions involving more than
two terms. 


%

%
%


\section{Discussion}
\label{discussion}

In \se{BHE}, \ref{CHE} and \ref{GHE}, solutions in terms of \PFQ functions
were derived for families of the Heun equations BHE, CHE and GHE. The
approach links linear equations with four regular singularites (and related
confluent cases) to linear equations with three regular singularities (and
related confluent cases), by linking both types of linear equations to the
canonical forms of the Abel AIR class of non-linear first order equations.
The link $AIR \leftrightarrow \PFQ$ is developed in \cite{abel3}, and the
link $AIR \leftrightarrow Heun$ is presented in \se{Theory}. This link also
provided a natural way to determine the special
Liouvillian cases of the Heun solutions to the BHE, CHE and GHE here
treated, and
permits studying Abel equation problems by reformulating them in terms of
linear equations.

In \se{alternative}, that approach is shown to be equivalent to performing
the non-local transformation \eq{MobiusR_0110} on the \PFQ equations
\eq{1F1_BHE}, \eq{1F1_CHE} and \eq{2F1_GHE}, leading to Heun equations with
two important properties: 1) further auxiliary equations which can be
derived from them for $p\equiv y'$ are of \PFQ type; 2) when written in
normal form, these Heun equations obtained using \eq{MobiusR_0110} are
identical to the BHE \eq{BHE_1}, CHE \eq{CHE_2} and GHE \eq{GHE_3} solved in
the sections previous to \se{alternative}. This approach leads in a simpler
manner to the same solutions \eq{tr_tot}, \eq{tr_tot_2} and \eq{tr_tot_3}, and also to the
equivalent forms of these solutions free of integrals,
\eq{tr_tot_1_free_of_integrals}, \eq{tr_tot_2_free_of_integrals} and
\eq{tr_tot_3_free_of_integrals}. All the solutions presented were verified
for correctness using symbolic computation software.

Besides the presentation in \se{BHE}, \ref{CHE} and \ref{GHE}, the existence
of a connection between the ``Heun and related confluent equations" on the
one hand and the ``AIR \eq{AIR_rho} and the different possible
multiplicities of its roots $\rho_i$" on the other hand, can be seen more
straightforwardly by transforming not \eq{can_forms} but \eq{AIR_rho} into a
linear equation\footnote{For that purpose, apply first \yx\ to \eq{AIR_rho}
at $\{s_2=1,r_2=-a\}$, then apply \eq{riccati_to_linear} to the resulting
Riccati equation.}, resulting in 

\be{AL_AIR}
\yt= \l(  \fr{1}{ x-a }
+ {\fr {R_{{3}}}{x-\rho_{{3}}}} + {\fr {R_{{2}}}{x-\rho_{{2}}}}+{\fr {R_{{1}}}{x-\rho_{{1}}}}\r) \y1
+ {\fr { \l( s_{{0}}x+r_{{0}} \r)  \l( a-x \r)}{ \l( x-\rho_{{3}} \r) ^{2} \l( x-\rho_{{2}}
 \r) ^{2} \l( x-\rho_{{1}} \r) ^{2}}}\, y
\ee

\ni This is a Heun equation with its four regular singularites at
$\{\rho_1,\rho_2,\rho_3,a\}$, where $R_1 + R_2 + R_3 = -3$,

\begin{equation}
R_{{2}}={\fr {{\rho_{{2}}}^{2}- \l( s_{{1}}+\rho_{{3}}+\rho_{{1}}
 \r) \rho_{{2}} + \rho_{{3}}\rho_{{1}} - r_{{1}}}{ \l( \rho_{{1}}-
\rho_{{2}} \r)  \l( \rho_{{2}}-\rho_{{3}} \r) }}
\ee

\ni and $\{R_1, R_3\}$ are obtained from $R_2$ multiplying by $-1$ and
respectively swapping $\rho_2 \leftrightarrow \rho_1$ and $\rho_2
\leftrightarrow \rho_3$. Through the confluence processes which coalesce
singularities in \eq{heun}, generating the CHE and BHE confluent equations, one
coalesces the corresponding singularities $\rho_i$ of \eq{AL_AIR}, generating
the same type of confluent equations, and that is equivalent to having
multiple roots $\rho_i$ in the AIR \eq{AIR_rho}. 

By rewriting \eq{AL_AIR} in normal form (see \eq{GHE_3}), the number of
irreducible parameters involved is shown to be four instead of six as in
\eq{GHE_NF}. That explains the restrictions on the Heun parameters of the
BHE, CHE and GHE families discussed in the previous sections. In the three
cases, one parameter is fixed and another one is dependent on those
remaining. 


Different from the BHE, CHE and GHE cases, in the case of the DHE
\eq{DHE_NF} and THE \eq{THE_NF} the approach considered in this paper does
not lead to new solutions. That can be seen by applying to \eq{AL_AIR} the
DHE and THE confluence processes \cite{decarreau}, in both cases arriving at
equations already of \PFQ type. That status of things is somewhat expected:
the AIR class is generated from the three canonical forms \eq{can_forms} and
these are, in their general form, already linked to GHE, CHE and BHE
families.


Independent of the possibility, developed here, of expressing the solutions
to the BHE \eq{BHE_1}, CHE \eq{CHE_2} and GHE \eq{GHE_3} normal forms {\em
without} introducing ``Heun functions", these functions have been developed
consistenly during the last years and will most certainly form part of the
standard mathematical language in the near future. That can be inferred from
the relevance of Heun equations in applications. In this framework, the
results of this paper could be seen as the identification of multi-parameter
special cases of Heun functions of the BHE, CHE and GHE types, respectively
admitting the integral representations \eq{tr_tot}, \eq{tr_tot_2} and
\eq{tr_tot_3}, and the linear combinations of \PFQ functions with
variable coefficients \eq{tr_tot_1_free_of_integrals},
\eq{tr_tot_2_free_of_integrals} and \eq{tr_tot_3_free_of_integrals}. The
mathematical properties and the relevance of these special cases in
applications require further investigation.

This link between Heun and \PFQ second order linear equations through Abel
non-linear equations of first order seems to be the simplest case of
a link between linear equations with N and N-1 singularities, through ``Abel
AIR like" equations, for which the numerator of the right-hand-side has degree
N-1 at most\footnote{For Abel equations of the first kind, $N = 3$; for Abel
equations of the second kind like \eq{abel_2k}, $N \leq 3$.}. For example,
if instead of \eq{AIR_rho} we depart from

\be{AIR_rho4}
\y1={\frac{(y-\rho_1)\, (y-\rho_2)\, (y-\rho_3)\, (y-\rho_4)}
{ \l( s_{2}\,x^2 + s_{1}\,x + s_{0} \r) y + r_{2}\,x^2 + r_{1}\,x + r_{0}}}
\ee

\ni that is, an equation with structure similar to the AIR \eq{AIR_rho} but
whose numerator of the right-hand-side is of degree four, then by applying
\yx\ to obtain a Riccati equation and transforming the latter into a
second order linear equation, we obtain an equation similar to \eq{AL_AIR} but with
five regular singular points,

\be{AL_AIR4}
\yt= \l(  \fr{1}{ x-a }
+ {\fr {R_{{4}}}{x-\rho_{{4}}}}
+ {\fr {R_{{3}}}{x-\rho_{{3}}}} + {\fr {R_{{2}}}{x-\rho_{{2}}}}+{\fr {R_{{1}}}{x-\rho_{{1}}}}\r) \y1
+ {\fr 
    { \l( s_{{0}}x+r_{{0}} \r)  \l( a-x \r)}
    { \l( x-\rho_{{4}} \r)^{2} \l( x-\rho_{{3}} \r)^{2} \l( x- \rho_{{2}}\r)^{2} \l( x-\rho_{{1}} \r)^{2}}}\, y,
\ee

\ni which, together with its confluent cases, can be linked through
\eq{AIR_rho4}, this time to the Heun equations, using the same
approach presented in the previous sections relating Heun to \PFQ equations.

Analogously, $n^{th}$ order ($n > 2$) linear equations in $y(x)$ can
also be reduced to ``Riccati like" non-linear equations of order $n-1$, due
to their invariance under scalings of $y$. It is therefore reasonable to
expect that a link equivalent to the one discussed in this work also
exists between linear equations with N and N-1 singularities in the $n^{th}$
($n > 2$) order case.



\section*{Appendix}
\label{Appendix_A}

In \se{BHE}, \ref{CHE} and \ref{GHE}, solutions were derived for the
equations in {\em normal form} BHE \eq{BHE_1}, CHE \eq{CHE_2} and GHE
\eq{GHE_3}. Given a second order linear ODE 

\begin{equation}
\yt + c_1 \y1 + c_0  y = 0
\ee

\ni where the $c_i \equiv c_i(x)$, the corresponding normal form, 

\begin{equation}
\yt + \l( c_{{0}} - ({c_{{1}}}^{2}+2\,{c_{1}}^{'})/4 \r) y=0
\ee

\ni is obtained
by changing $ y \rightarrow {\exp(-{\int \!c_{1}\, dx/2})}\,y$.

Regarding Heun equations, one advantage of the normal form is that the
general or confluent type of the equation is evident in the partial fraction
decomposition of the coefficient of $y$ (see eqs. \eq{GHE_NF} to
\eq{THE_NF}). Also, two different equations related by $y \rightarrow
P(x)\,y$ have the same normal form and recognizing this equivalence is
relevant for computational purposes. On the other hand, for different
reasons, special functions are frequently defined as solutions to equations
in {\em canonical form}. This appendix relates the normal and canonical
forms of the BHE, CHE and GHE, expressed in terms of irreducible
parameters\footnote{A few of the equations shown in the paper are repeated
here for ease of reading.}, following the notation of \cite{decarreau}, thus
permitting a simple translation of the results presented. 

\subsubsection*{The Biconfluent Heun equation}

The BHE canonical form is given in terms of four constant parameters
$\{\alpha,\beta,\gamma,\delta\}$ by


\be{BHE_can}
\yt+ \l( {\fr {1+\alpha}{x}}-\beta-2\,x \r) \y1+ \l( {\gamma}-\alpha-2 - {\fr {\delta+ \l( 1+\alpha \r) \beta}{2\,x}
} \r) y=0;
\ee

\ni The BHE in normal form \eq{BHE_NF} in terms of four parameters $\{B,C,D,E\}$ is


\be{BHE_NF_Appendix}
\yt+ \l( -{x}^{2}+B\,x+C+{\fr {D}{x}}+{\fr {E}{{x}^{2}}} \r) y  =
0
\ee

\ni The BHE normal form \eq{BHE_1} solved in \se{BHE}, there written in terms of two parameters $\{\sigma,\tau\}$, is

\be{BHE_P}
\yt - \l( {x}^{2}+2\,\sigma\,x+{\tau}^{2} + {\fr {\tau}{x}} +
\fr{3}{4\,x^{2}} \r) y = 0
\ee

\ni The parameters $\{B,C,D,E\}$ in \eq{BHE_NF_Appendix} are related to
$\{\sigma,\tau\}$ by

\begin{equation}
B=-2\,\sigma,
\ \ \ 
C=-{D}^{2},
\ \ \ 
D=-\tau,
\ \ \ 
E=-3/4
\ee

\ni The parameters $\{\alpha,\beta,\gamma,\delta\}$ in \eq{BHE_can} are related to
$\{B,C,D,E\}$ by

\begin{equation}
{\alpha}^{2}=-4\,E+1,
\ \ \ 
\beta=-B,
\ \ \ 
\gamma={B}^{2}/4+C,
\ \ \ 
\delta=-2\,D
\ee

\ni So the parameters $\{\alpha,\beta,\gamma,\delta\}$ are related to $\{\sigma,\tau\}$ by

\begin{equation}
{\alpha}^{2}=4,
\ \ \ 
\beta=2\,\sigma,
\ \ \ 
\gamma={\sigma}^{2}-{\tau}^{2},
\ \ \ 
\delta=2\,\tau
\ee


\ni At these values of $\{\alpha,\beta,\gamma,\delta\}$, for $\sigma = \pm
\tau$, \eq{BHE_can} admits Liovillian solutions and for $\sigma^2 \neq
\tau^2$ the solution is obtained from \eq{tr_tot_1_free_of_integrals}.

\subsubsection*{The Confluent Heun equation}

The CHE canonical form is given in terms of five constant parameters
$\{\alpha,\beta,\gamma,\delta,\eta\}$ by


\be{CHE_can}
\yt+ \l( \alpha+{\fr {\beta+1}{x}}+{\fr {\gamma-1}{x-1}}\r) \y1
+ {\fr 
    {\l( 2\,\delta + \alpha \l( \beta+\gamma+2 \r)  \r) x
        + 2\,\eta
        + \beta + \l( \gamma-\alpha \r)  \l( \beta+1 \r)}
    {2\,x \l( x-1 \r) }\,y}=0
\ee

\ni The CHE in normal form \eq{CHE_NF} in terms of five parameters $\{A,B,C,D,E\}$ is


\be{CHE_NF_Appendix}
\yt+ \l( A+{\fr {B}{x}}+{\fr {C}{x-1}}+{\fr {D}{{x}^{2}}}+{
\fr {E}{ \l( x-1 \r) ^{2}}} \r) y = 0
\ee

\ni The CHE normal form \eq{CHE_2}  solved in \se{CHE}, there written in terms
of three parameters $\{\lambda,\sigma,\tau\}$, is

\be{CHE_P}
\yt - 
\l( 
    {\lambda}^{2} 
    + {\fr {2\,( \sigma - 1) {\lambda}^{2} - \tau\,\lambda + 1/2}{x}}
    + {\fr {\tau \lambda - 1/2}{x-1}}
    + {\fr {( {\tau}^{2} -2\,\sigma + 1 )\,{\lambda}^{2} - 1/4}{{x}^{2}}}
    + \fr{3}{4 \l( x-1 \r)^{2}}
    \r) y = 0
\ee

\ni The parameters $\{A,B,C,D,E\}$ in \eq{CHE_NF_Appendix} are related to
$\{\lambda,\sigma,\tau\}$ by

\begin{equation}
A =-{\lambda}^{2},
\ \ \ 
B = 2\l( 1-\sigma \r) {\lambda}^{2} + \tau\,\lambda -\fr{1}{2},
\ \ \ 
C = \fr{1}{2} - \tau\,\lambda,
\ \ \ 
D = \fr{1}{4} + \l( 2\,\sigma-{\tau}^{2} -1 \r) {\lambda}^{2},
\ \ \ 
E = -3/4,
\ee

\ni The parameters $\{\alpha,\beta,\gamma,\delta,\eta\}$ in \eq{CHE_can} are related to
$\{A,B,C,D,E\}$ by

\begin{equation}
{\alpha}^{2}=-4\,A,
\ \ \ 
{\beta}^{2}=-4\,D+1,
\ \ \ 
{\gamma}^{2} = 4\,\gamma - 4\,E  - 3
\ \ \ 
\delta=C+B-\alpha,
\ \ \ 
\eta=-\fr{1}{2}-B-\beta
\ee

\ni So the relation between $\{\alpha,\beta,\gamma,\delta,\eta\}$ and
$\{\lambda,\sigma,\tau\}$ is

\begin{equation}
{\alpha}^{2}=4\,{\lambda}^{2},
\ \ \ 
{\beta}^{2}=4 \l( 1-2\,\sigma+{\tau}^{2} \r) {\lambda}^{2}
\ \ \ 
{\gamma}^{2}=4\,\gamma,
\ \ \ 
\delta= 2\l( 1 - \sigma \r) {\lambda}^{2}-\alpha,
\ \ \ 
\eta= 2 \l( \sigma-1 \r) {\lambda}^{2}-\tau\,\lambda-\beta
\ee

\ni At these values of $\{\alpha,\beta,\gamma,\delta,\eta\}$, the CHE
\eq{CHE_can} admits Liouvillian solutions for $\sigma = \pm \tau$, and for
$\sigma^2 \neq \tau^2$ the solution is obtained from \eq{tr_tot_2_free_of_integrals}.

\subsubsection*{The General Heun equation}

The GHE canonical form is written in terms of seven
constant parameters $\{\alpha,\beta,\gamma,\delta,\epsilon,a,q\}$ as 


\be{GHE_can}
\yt+ \l( {\fr {\gamma}{x}}+{\fr {\delta}{x-1}}+{\fr {\epsilon}{x-a}} \r) \y1+{\fr {\alpha\,\beta\, x-q }
{x \l( x-1 \r)  \l( x-a \r) }\,y}=0
\ee

\ni where $\gamma+\delta+ {\epsilon}=\alpha+\beta+1$ and $a\neq 0,a\neq 1$.
In \cite{decarreau}, the numerator of the coefficient of $y$ of this
equation is written as ${\alpha\,\beta \l( x-h \r) }$. The notation in
\eq{GHE_can} is the one used in \cite{ronveaux}, was apparently first
adopted in \cite{erdelyi}, and has the advantage that one can take $\alpha =
0$ (or $\beta = 0$) without eliminating the term in $y$
completely\footnote{The differences in notation with respect to
\cite{decarreau} are the coefficient of $y$ and the use of
$\{\gamma,\delta,\epsilon,\alpha,\beta,q\}$ in place of
$\{\alpha,\beta,\gamma,\delta,\eta,h\}$.}. The GHE in normal form
\eq{GHE_NF} in terms of six parameters $\{a,A,B,D,E,F\}$ is


\be{GHE_NF_Appendix}
\yt+ \l( {\fr {A}{x}} + {\fr {B}{x-1}} -{\fr {A+B}{x-a}} 
+ {\fr {D}{{x}^{2}}} + {\fr {E}{ \l( x-1 \r) ^{2}}} + {\fr {F}{
\l( x-a \r) ^{2}}} \r) y = 0
\ee

\ni The GHE normal form solved in \se{GHE} (see \eq{GHE_3}), written in terms of four parameters $\{a,\sigma,\tau,\Delta\}$, is


\begin{eqnarray}
\label{GHE_P}
\lefteqn{ \yt = 
\l( 
{\fr 
    {2\,{a}^{2} \l( a-1 \r) {\Delta}^{2} - 2\sigma\,a \l( 2\,a -1\r) \Delta +  \l( 2\,{\tau}^{2} -1/2\r) a + \tau + 1/2} 
    {a\,x}}
\r.
}
& &
\\*[0.1 in]
\nonumber
& &
- {\fr 
    {2\l( a \l( a-1 \r)^{2}{\Delta}^{2} - \sigma \l( 2\,a -1\r)  \l( a-1 \r) \Delta + \l( \tau - 1/2\r) \l( \l( \tau + 1/2 \r) a - \tau \r)\r) }
    {\l( a-1 \r) \l( x-1 \r) }}
+ {\fr 
    {\tau -a + 1/2 }
    { a\l( a-1 \r) \l( x-a \r)}}
\\*[0.1 in]
\nonumber
& &
\l.
+ {\fr 
    {{a}^{2}{\Delta}^{2} - 2\,a\,\sigma\,\Delta + {\tau}^{2} - 1/4}
    {{x}^{2}}}
+ {\fr 
    {\l( a-1 \r) ^{2}{\Delta}^{2} - 2\,\sigma \l( a-1 \r) \Delta + {\tau}^{2} - 1/4}
    {\l( x-1 \r) ^{2}}} 
+ \fr{3}{4 \l( x-a \r) ^{2}}
\r) y
\end{eqnarray}

\ni The parameters $\{A,B,D,E,F\}$ in \eq{GHE_NF_Appendix} are related to
$\{a,\sigma,\tau,\Delta\}$ by

\begin{eqnarray}
\nonumber
A & = & 
-2\,a \l( a-1 \r) {\Delta}^{2} + 2\l( 2\,a - 1 \r) \sigma\,\Delta 
- 2\,{\tau}^{2} - {\fr {\tau+1/2}{a}} + \fr{1}{2},
\\
\nonumber
B & = & 
2\,a \l( a-1 \r) {\Delta}^{2} - 2\l( 2\,a-1\r) \sigma\,\Delta
+ 2\,{\tau}^{2}+{\fr {\tau-a/2}{a-1}},
\\
D & = & -{a}^{2}{\Delta}^{2} + 2\,a\,\sigma\,\Delta -{\tau}^{2} + 1/4,
\\
\nonumber
E & = & - \l( a-1 \r) ^{2}{\Delta}^{2} + 2 \l( a-1\r) \sigma\,\Delta - {\tau}^{2} + 1/4,
\\
\nonumber
F & = & -3/4
\end{eqnarray}

\ni The parameters $\{\alpha,\beta,\gamma,\delta,\epsilon,q\}$ in \eq{GHE_can} are related to
$\{a,A,B,D,E,F\}$ by

\begin{eqnarray}
\nonumber
{\gamma}^{2} & = & -4\,D+2\,\gamma,
\ \ \ \ \ {\delta}^{2}  =  -4\,E+2\,\delta,
\ \ \ \ \ {\epsilon}^{2}  =  -4\,F+2\,\epsilon,
\\
\nonumber
{\alpha}^{2}& = & \l( \delta + \epsilon + \gamma -1 \r) \alpha -  \fr{\l( \gamma +\delta \r)\epsilon}{2}- \fr{\gamma\,\delta}{2} + \l( a-1 \r) B + a\,A
\\
\nonumber
\beta & = & \gamma+\delta+ {\epsilon}-\alpha-1,
\ \ \ \ \ q  =  \fr{\l( a\,\delta + \epsilon \r) \gamma}{2}-a\,A
\end{eqnarray}

\ni So the parameters $\{\alpha,\beta,\gamma,\delta,\epsilon, q\}$ are related to $\{a,\sigma,\tau,\Delta\}$ by

\begin{eqnarray}
\nonumber
{\gamma}^{2} & = & 2\,(2\,{a}^{2}{\Delta}^{2} - 4\,a\,\sigma\,\Delta + 2\,{\tau}^{2} + \gamma) -1
\\
\nonumber
{\delta}^{2} & = & 2\,(2 \l( a-1 \r)^{2}{\Delta}^{2} + 4\,\sigma\l( 1-a \r) \Delta + \delta + 2\,{\tau}^{2}) - 1
\\
{{\epsilon}}^{2} & = & 2\,\epsilon +3
\\
\nonumber
{\alpha}^{2} & = & 2\,a\l( 1-{a} \r) {\Delta}^{2} + 2 \l( 2\,a-1 \r)
\sigma\,\Delta + \l( \gamma + \delta + {\epsilon} -1 \r) \alpha - 2\,{\tau}^{2} - (\gamma\,\delta + 1 + \l(\gamma+\delta \r) \epsilon)/2
\\
\nonumber
\beta & = & \gamma+\delta+ {\epsilon}-\alpha-1
\\
\nonumber
q & = & 2\,{a}\,\Delta \l( a\l( a-1 \r) {\Delta} - \sigma \l( 2\,a -1\r) \r) + \fr {a\l( \gamma\,\delta + 4\,{\tau}^{2}-1 \r) }{2} + \fr{\gamma\,\epsilon}{2} + \tau + \fr{1}{2}
\end{eqnarray}

\ni At these values of $\{\alpha,\beta,\gamma,\delta,\epsilon,q\}$, for $\sigma
= \pm \tau$ the GHE \eq{GHE_can} admits Liouvillian solutions, and for
$\sigma^2 \neq \tau^2$ the solution is obtained from \eq{tr_tot_3_free_of_integrals}.

\subsubsection*{Verifying solutions using symbolic computation}
\label{odetest}

In presentations like this one, where equations and solutions involving many
parameters and non trivial special functions are involved, it is of use to
be able to verify the correctness of the solutions derived, in some way
alternative to the one presented. For that purpose, the input lines, written
in the Maple symbolic computation syntax, for the BHE \eq{BHE_1}, the CHE
\eq{CHE_2} and the GHE \eq{GHE_3} equations and their respective solutions
\eq{tr_tot_1_free_of_integrals}, \eq{tr_tot_2_free_of_integrals} and
\eq{tr_tot_3_free_of_integrals}, are given, so that they can be copied from the
online version of this paper.


The 2-parameter BHE \eq{BHE_1} is written in Maple syntax as

\begin{verbatim}
    > BHE := diff(y(x),x,x) - (x^2 + 2*sigma*x + tau^2 + tau/x + 3/4/x^2)*y(x) = 0;
\end{verbatim}

\ni and its solution \eq{tr_tot_1_free_of_integrals} is written as

\begin{verbatim}
    > BHE_sol := y = exp(-sigma*x-1/2*x^2)/x^(1/2)/(x+sigma)*((Lambda*KummerU(1/4*tau^2
    > - 1/4*sigma^2,1/2,(x+sigma)^2)-4*KummerU(1/4*tau^2-1/4*sigma^2-1,1/2,(x+sigma)^2))
    > * _C1+((tau^2-sigma^2-2)*KummerM(1/4*tau^2-1/4*sigma^2-1,1/2,(x+sigma)^2)-Lambda
    > * KummerM(1/4*tau^2-1/4*sigma^2,1/2,(x+sigma)^2))*_C2);
    > Lambda := sigma^2+tau^2+4*x^2+2*(3*sigma-tau)*x-2*sigma*tau-2;
\end{verbatim}

\ni After entering these lines in a Maple session, to verify this solution one
can use the Maple {\bf odetest} command, as in \verb-> odetest( BHE_sol, BHE );-which returns zero, confirming that the
solution cancels the equation.
%
%
The 3-parameter CHE \eq{CHE_2} is written in Maple syntax as

\begin{verbatim}
    > CHE := diff(y(x),x,x) = ((-1+2*tau*lambda)/(2*x-2)+1/2*(1+(-4+4*sigma)
    > * lambda^2-2*tau*lambda)/x+1/4*(-1+(4-8*sigma+4*tau^2)*lambda^2)/x^2
    > + 3/4/(x-1)^2+lambda^2)*y(x);
\end{verbatim}

\ni and its solution \eq{tr_tot_2_free_of_integrals} is written as

\begin{verbatim}
    > CHE_sol := y(x) = 1/(x-1)^(1/2)*(((tau+sigma)*lambda*WhittakerM(mu,nu,2*lambda*x)
    > + ((-sigma+1)*lambda-nu)*WhittakerM(-1+mu,nu,2*lambda*x))*_C1+(lambda*(tau-sigma)
    > * WhittakerW(-1+mu,nu,2*lambda*x)+WhittakerW(mu,nu,2*lambda*x))*_C2);
    > mu := 1/2-lambda*sigma+lambda; nu := lambda*(-2*sigma+1+tau^2)^(1/2);
\end{verbatim}



\ni The 4-parameter GHE \eq{GHE_3} is written in Maple syntax as

\begin{verbatim}
    > GHE := diff(y(x),x,x) = ((2*a^2*(a-1)*Delta^2-2*sigma*a*(2*a-1)*Delta
    > + (2*tau^2-1/2)*a+tau+1/2)/x/a-2*(a*(a-1)^2*Delta^2-sigma*(2*a-1)*(a-1)*Delta
    > + (tau-1/2)*((tau+1/2)*a-tau))/(x-1)/(a-1)+(tau-a+1/2)/a/(a-1)/(x-a)
    > + (Delta^2*a^2-2*a*sigma*Delta+tau^2-1/4)/x^2+((a-1)^2*Delta^2-2*(a-1)*sigma
    > * Delta+tau^2-1/4)/(x-1)^2+3/4/(x-a)^2)*y(x);
\end{verbatim}

\ni and its solution \eq{tr_tot_3_free_of_integrals} is written as

\begin{verbatim}
    > GHE_sol := y(x) = 1/(x-a)^(1/2)*(x-1)^(1/2+Sigma)*((1/2*(Tau-Sigma-Delta)
    > * (-Sigma+Delta-1+Tau)*(x^(5/2-Tau)-x^(3/2-Tau))*hypergeom([Sigma-Delta+2
    > - Tau, Sigma+Delta-Tau+1],[2-2*Tau],x)+hypergeom([Sigma+Delta-Tau, Sigma-Delta
    > + 1-Tau],[1-2*Tau],x)*((Tau-Sigma-Delta)*x^(3/2-Tau)+x^(1/2-Tau)
    > * (-Tau+a*Delta+tau))*(-1/2+Tau))*_C1+_C2*(1/2*(Sigma+Delta+Tau)*(Sigma-Delta
    > + 1+Tau)*(-x^(5/2+Tau)+x^(3/2+Tau))*hypergeom([Sigma-Delta+2+Tau, Sigma
    > + Delta+Tau+1],[2+2*Tau],x)+(1/2+Tau)*((-Tau-Sigma-Delta)*x^(3/2+Tau)
    > + x^(1/2+Tau)*(Tau+a*Delta+tau))
    > * hypergeom([Sigma+Delta+Tau, Sigma-Delta+1+Tau],[1+2*Tau],x)));
    > Sigma := sqrt((a-1)^2*Delta^2-2*(a-1)*sigma*Delta+tau^2);
    > Tau := sqrt(a^2*Delta^2-2*a*sigma*Delta+tau^2);
\end{verbatim}



\medskip
\ni {\bf Acknowledgments}
\medskip

\noindent This work was supported by the MITACS NCE, the Centre of
Experimental and Constructive Mathematics of Simon Fraser University and the
Maplesoft division of Waterloo Maple Inc. The author thanks K.
von B\"ulow for a careful reading of this paper, as well as one of the
referees for her/his fruitful comments.


\end{document}